\documentclass[]{interact}
\usepackage{mathrsfs}
\usepackage{epstopdf}
\usepackage[caption=false]{subfig}
\usepackage{algorithm}
\usepackage{multirow,multicol}
\usepackage{accents}
\usepackage[numbers,sort&compress]{natbib}
\bibpunct[, ]{[}{]}{,}{n}{,}{,}
\renewcommand\bibfont{\fontsize{10}{12}\selectfont}

\theoremstyle{plain}

\newtheorem{res}{Result}
\theoremstyle{definition}

\theoremstyle{remark}

\begin{document}

\articletype{ARTICLE TEMPLATE}

\title{Robust Bayesian inference for nondestructive one-shot device testing data under competing risk using Hamiltonian Monte Carlo method}

\author{
\name{Shanya Baghel\textsuperscript{a} and Shuvashree Mondal\textsuperscript{a}\thanks{CONTACT Shuvashree Mondal. Email: shuvasri29@iitism.ac.in} }
\affil{\textsuperscript{a}Department of Mathematics and Computing, IIT (Indian School of Mines) Dhanbad, Jharkhand, India.}
}

\maketitle

\begin{abstract}
The prevalence of one-shot devices is quite prolific in engineering and medical domains.  Unlike typical one-shot devices, nondestructive one-shot devices (NOSD) may survive multiple tests and offer additional data for reliability estimation.  This study aims to implement the Bayesian approach of the lifetime prognosis of NOSD when failures are subject to multiple risks.  With small deviations from the assumed model conditions, conventional likelihood-based Bayesian estimation may result in misleading statistical inference, raising the need for a robust Bayesian method.  This work develops Bayesian estimation by exploiting a robustified posterior based on the density power divergence measure for NOSD test data.  Further, the testing of the hypothesis is carried out by applying a proposed Bayes factor derived from the robustified posterior.  A flexible Hamiltonian Monte Carlo approach is applied to generate posterior samples.  Additionally, we assess the extent of resistance of the proposed methods to small deviations from the assumed model conditions by applying the influence function (IF) approach.  In testing of hypothesis, IF reflects how outliers impact the decision-making through Bayes factor under null hypothesis.  Finally, this analytical development is validated through a simulation study and a data analysis based on cancer data.
\end{abstract}


\section{Introduction}
The nondestructive one-shot devices (NOSD) are widely prevalent in engineering and medical domains.  Unlike one-shot devices, all NOSD units may not be destroyed after testing.  The surviving NOSD units can continue the trials, offering additional data for reliability estimation.  Devices like metal fatigue, spare wheels, light bulbs, electric motors, etc., are NOSD.  Recently, the reliability analysis of NOSD has garnered the attention of various studies \cite{balakrishnan2022robust,balakrishnan2022non,balakrishnan2022restricted,balakrishnan2023robust,balakrishnan2024non}.  Further, one can witness many real-life scenarios where the failure of NOSD may be subject to many potential risks.  For example, a light bulb can fail due to overheating, overloading, or high vibration.  The competing risk analysis focuses on identifying the cause of the ultimate failure.  In the literature, a significant amount of work is devoted to study competing risk analysis \cite{bet2015,bnsl2015,balakrishnan2023power,balakrishnan2024robust}, although competing risk analysis for NOSD is barely explored.

Existing literature on NOSD reliability predominately employs exponential, Weibull, and lognormal as lifetime distributions.  To capture a broad spectrum of hazard functions,  this study models the lifetime of NOSD under independent competing risks with a two-parameter Lindley distribution.  Introduced by Lindley \cite{lind1958, lindley1970introduction} as lifetime distribution, Ghitany et al. \cite{ghit2008} demonstrated that the Lindley distribution offers improved modelling capabilities over exponential distribution regarding mathematical properties like mode, coefficient of variation, skewness, kurtosis and hazard rate shape.  Mazucheli and Achcar \cite{m2011} also recommended it as an alternative to exponential and Weibull distributions in competing risk scenarios.  While Castilla Gonzalez \cite{castilla2021robust} employed one-parameter Lindley distribution in destructive one-shot device testing data, analysis in a competing risk setup under two-parameter Lindley lifetime distribution for NOSD is unprecedented.

Bayesian inference is quite essential when some prior information about the model parameters is available.  However, conventional Bayesian estimation method \cite{chandra2024inference,sun2024bayesian} incorporating likelihood-based posteriors can be unreliable with contaminated data.  Ghosh and Basu \cite{g2016} addressed it by introducing robust Bayes estimation, where the density power divergence (DPD) measure \cite{basu1998} has substituted the likelihood function in the posterior density function.  To the best of the authors' knowledge, robust Bayesian inference in the context of NOSD is yet to be studied, which brings novelty to this study.  

This study adopts Normal and Dirichlet distributions \cite{f2009} as priors in the robust Bayesian framework of NOSD test data.  While Gibbs sampler and Metropolis-Hastings algorithms are frequently used for posterior estimation, they may be inefficient in exploring the target distribution with high dimensional or highly correlated variables \cite{thach2019reparameterized}.  Hamiltonian Monte Carlo (HMC), introduced by Neal \cite{neal2011mcmc,neal2012bayesian} to the application of statistics, offers a solution, providing accurate results and flexibility in complex models \cite{thanh2021additive,abba2023new}.  For an in-depth explanation of HMC, one can refer to \cite{monnahan2017faster,abba2024robust} and the references therein.  The present study is the first attempt to seek HMC to solve the robust Bayes estimation problem of NOSD test data under competing risk.  

Another critical aspect of the Bayes framework is the testing of hypotheses through the Bayes factor, initially introduced by Jeffreys \cite{jeffreys1973scientific, jeffreys1935some, jeffreys1998theory} and subsequently applied by numerous researchers.  Further, influence function analysis is evident in the study of robustness.  Basu et al. \cite{basu1998} and Ghosh and Basu \cite{g2016} derived influence functions for robust Bayes and DPD-based estimates, respectively. However, the influence function analysis of the robust Bayes factor evaded the attention of researchers and its application for NOSD test data is yet to be conducted.

The present article develops a robust Bayesian estimation method for reliability analysis of NOSD testing data under accelerated life testing with independent competing risks and interval monitoring, modelled with a two-parameter Lindley lifetime distribution. The estimation procedure relies on a weighted robust Bayes estimation (WRBE)\cite{g2016} method, creating a robustified posterior density through the exponential form of the maximizer equation using the DPD measure.  Additionally, this study explores the robust testing of the hypothesis by applying a Bayes factor derived from the robustified posterior.  Furthermore, the influence functions are derived and examined thoroughly to assess the robust behaviour of the point estimators in the Bayesian framework.  In the testing of hypotheses, the influence function reflects how outlying observation can influence the Bayes factor under the null hypothesis, potentially affecting decision-making.

The rest of the article comprises sections divided as follows.  Section \ref{sec2} is concerned with competing risk model building.  Section \ref{sec3} introduces the robust Bayes estimation method based on density power divergence.  Testing of the hypothesis based on the robust Bayes factor is developed in section \ref{sec4}.  In section \ref{sec5}, we study the robustness properties of the estimators through the influence function.  Section \ref{sec6} deals with the numerical study of the theoretical results developed in previous sections through a simulation study and data analysis.  Finally, concluding remarks are given in Section \ref{sec7}.

\section{ Model Description}\label{sec2}
Nondestructive one-shot devices (NOSD) are studied under accelerated life testing (ALT), where $G$ devices are distributed into $I$ independent groups.  Let the number of devices in $ith$ group be $g_i$ where $G=\sum_{i=1}^{I}g_i$ and $J$ type of stress factors be implemented on these devices quantified by $s_{ij}$; $j=1,\dots, J$, where $s_{i0}=1$ denotes normal operating condition.  The failure mechanism is subject to R-independent competing causes.  Let, $n_{ilr}$ denote the number of failures in the $ith$ group due to cause $r$ for $r=1,2,\dots,R,$ between the inspection times $(\tau_{i(l-1)},\tau_{il}]$  where $l=1,2,\dots,L$, and $\tau_{i0}=0.$   Therefore, the number of devices that survive after inspection time $\tau_{iL}$ for the $ith$ group is $k_i=g_i-\sum_{l=1}^{L}\sum_{r=1}^{R}n_{ilr}$.  Table \ref{tab1} shows the layout in tabular form.
\begin{table}[htb!]
\tbl{Model layout.}
{\begin{tabular}{lccccccccc} \toprule
{Groups}&{Devices} & \multicolumn{4}{c}{{Co-factors}} & \multicolumn{4}{c}{{Inspection Times}} \\ \cmidrule{3-6}
& & {Stress 1} & {Stress 2} & \dots & {Stress J} &\multicolumn{4}{c}{{}}\\ \midrule
1&$g_1$  &$s_{11}$ &$s_{12}$ & \dots &$s_{1J}$&$\tau_{11}$ &$\tau_{12}$ &\dots &$\tau_{1L}$ \\
2&$g_2$  &$s_{21}$ & $s_{22}$&\dots &$s_{2J}$&$\tau_{21}$ &$\tau_{22}$ &\dots &$\tau_{2L}$ \\
\vdots&\vdots &\vdots &\vdots &$\ddots$ & \vdots&\vdots & \vdots& $\ddots$&\vdots \\
i&$g_i$ &$s_{i1}$ &$s_{i2}$ &\dots &$s_{iJ}$ &$\tau_{i1}$ &$\tau_{i2}$ &\dots &$\tau_{iL}$ \\
\vdots&\vdots &\vdots &\vdots &$\ddots$ &\vdots &\vdots & \vdots& \vdots&\vdots \\
I&$g_I$  &$s_{I1}$ & $s_{I2}$&\dots &$s_{IJ}$&$\tau_{I1}$ &$\tau_{I2}$ &\dots &$\tau_{IL}$ \\ \midrule
{Groups}&{Survivals}& \multicolumn{8}{c}{{Failures}}\\\hline
1 &$k_1$ & \multicolumn{3}{c}{$(n_{111},\dots,n_{11R})$}  &\multicolumn{1}{c}{\dots}  &\multicolumn{4}{c}{$(n_{1L1},\dots,n_{1LR})$} \\
2 &$k_2$ & \multicolumn{3}{c}{$(n_{211},\dots,n_{21R})$}  &\multicolumn{1}{c}{\dots}  &\multicolumn{4}{c}{$(n_{2L1},\dots,n_{2LR})$} \\
\vdots &\vdots & \multicolumn{3}{c}{\vdots}  &\multicolumn{1}{c}{$\ddots$}  &\multicolumn{4}{c}{\vdots} \\
i &$k_i$ & \multicolumn{3}{c}{$(n_{i11},\dots,n_{i1R})$}  &\multicolumn{1}{c}{\dots}  &\multicolumn{4}{c}{$(n_{iL1},\dots,n_{iLR})$} \\
\vdots &\vdots & \multicolumn{3}{c}{\vdots}  &\multicolumn{1}{c}{$\ddots$}  &\multicolumn{4}{c}{\vdots} \\
I &$k_I$ & \multicolumn{3}{c}{$(n_{I11},\dots,n_{I1R})$}  &\multicolumn{1}{c}{\dots}  &\multicolumn{4}{c}{$(n_{IL1},\dots,n_{ILR})$} \\ \bottomrule
\end{tabular}}
\label{tab1}
\end{table}
The failure time of any device due to $r$th competing cause from the $i$th group for $i=1,2,\dots,I,$ $r=1,\ldots,R,$ is denoted by the random variable $T_{ir},$ which is assumed to follow a two-parameter Lindley distribution with shape parameter $\alpha_{ir}$ and scale parameter $\theta_{ir}$.  The cumulative distribution function and probability density function of $T_{ir}$ are as follows.
\begin{align*}
F_{ir}(t)&=1-\left(\frac{1+\alpha_{ir}\theta_{ir}+\theta_{ir} t}{\alpha_{ir}\theta_{ir}+1}\right)e^{-\theta_{ir} t}\;,\;f_{ir}(t)=\frac{\theta_{ir}^2}{\alpha_{ir}\theta_{ir}+1}\left(\alpha_{ir}+t\right)e^{-\theta_{ir} t},
\end{align*}
where, $(t,\theta_{ir})>0,\alpha_{ir}+\theta_{ir}>0.$ Both shape and scale parameters are related to stress factors in log-linear form as
\begin{equation*}
\alpha_{ir}=exp\left\{\sum_{j=0}^{J}a_{rj} s_{ij}\right\},\;
\theta_{ir}=exp\left\{\sum_{j=0}^{J}b_{rj} s_{ij}\right\}, 
\end{equation*}
where, $r=1,2,\dots,R.$ For the computational conveniences, two competing causes and one stress factor are considered without the additive constant and denote $s_{i1}=s_i.$   Hence, $\bm{\Lambda}=\{( a_{1},b_{1}, a_{2}, b_{2} )\}^{'}$ are the model parameters to be estimated.  The failure probabilities $p_{il1}$, $p_{il2}$ due to cause 1 and cause 2, respectively, in the interval $(\tau_{i(l-1)}, \tau_{il} )$ for $l=1,\ldots, L$ and the survival probability $  p_{i0}$ are obtained as follows.
\begin{align*}
p_{il1} &=P(\tau_{i(l-1)}<T_{i1}\leq\tau_{il},T_{i2}>T_{i1}),\; p_{il2}=P(\tau_{i(l-1)}<T_{i2}\leq\tau_{il},T_{i1}>T_{i2}),\\
p_{i0}&=P(T_{i1}>\tau_{iL},T_{i2}>\tau_{iL}).
\end{align*}
\begin{res}\label{res1}
Under the assumption of Lindley lifetime distribution, the failure and survival probabilities are derived as
 \begin{equation}
   \begin{aligned}
		p_{il1}&{=}\frac{\theta_{i1}^2\theta_i^{-3}A_{i}(\alpha_{i1},\theta_{i2})}{\prod_{r=1}^{2}(\alpha_{ir}\theta_{ir}+1)}\;;\;
		p_{il2}{=}\frac{\theta_{i2}^2\theta_i^{-3}A_{i}(\alpha_{i2},\theta_{i1})}{\prod_{r=1}^{2}(\alpha_{ir}\theta_{ir}+1)}\;;\;p_{i0}{=}\prod_{r=1}^{2}\left(\frac{1{+}\alpha_{ir}\theta_{ir}{+}\tau_{iL}\theta_{ir}}{\alpha_{ir}\theta_{ir}{+}1}\right)e^{-\theta_{ir}\tau_{iL}}.\label{eq1}
\end{aligned}  
 \end{equation}
 where,
 \begin{align*}
A_i(\alpha,\theta)&=e^{-\theta_i\tau_{i(l-1)}}\big\{A_{1i}(\tau_{i(l-1)},\theta)+A_{2i}(\alpha,\theta)\big\}-e^{-\theta_i\tau_{il}}\big\{A_{1i}(\tau_{il},\theta)+A_{2i}(\alpha,\theta)\big\}\,,\\
A_{1i}(\tau,\theta)&=\theta_i\tau\big\{\theta_i[\theta(\alpha_i{+}\tau){+}1]{+}2\theta\big\}\;;\;A_{2i}(\alpha,\theta)=\theta_i\big\{\theta(\alpha_i{+}\theta_i\alpha_{i1}\alpha_{i2}){+}1{+}\theta_i\alpha\big\}{+}2\theta\,,\\
\alpha_i&=\alpha_{i1}+\alpha_{i2}\;;\;\theta_i=\theta_{i1}+\theta_{i2}.
 \end{align*}
\end{res}
\noindent The log-likelihood function based on observed failure count data is obtained by
\begin{equation}
	ln\,L(\bm{\bm{\Lambda}})\propto\sum_{i=1}^{I}\left[k_i\, ln\,(p_{i0})]+\sum_{l=1}^{L}\sum_{r=1}^{2}n_{ilr}\, ln\,(p_{ilr})\right] . \label{eq2}
\end{equation}
Hence, MLE of $\bm{\bm{\Lambda}},$ denoted by $\;\hat{\bm{\bm{\Lambda}}}=\{\hat{a}_1,\hat{b}_1,\hat{a}_2, \hat{b}_2\}$ would be derived as 
\begin{equation}
\hat{\bm{\bm{\Lambda}}}=arg\mathop{max}_{\bm{\bm{\Lambda}}}\,ln L(\bm{\bm{\Lambda}}) . \label{eq3}
\end{equation}
Provided $\sum_{i=1}^{I}\sum_{l=1}^{L}\sum_{r=1}^{2}n_{ilr}>0.$ However, MLE cannot provide a valid estimated value when data comes with outliers.  Therefore, some robust estimation method needs to be developed.  This study discusses a robust estimation method for NOSD based on the density power divergence (DPD) measure proposed by Basu et al. \cite{basu1998}.  

\section{Robust Bayes Method of Estimation}\label{sec3}
Bayesian inference is of paramount interest when some prior information about the model parameters is available.  The major drawback with conventional Bayes estimation based on likelihood-based posterior is that it may not produce a good estimated value when data comes with contamination. Ghosh and Basu \cite{g2016} proposed to solve the non-robustness problem by replacing the likelihood function in the posterior with the Density power divergence (DPD) based loss function, where the derived posterior is called a pseudo posterior.  For NOSD data, the DPD measure \cite{basu1998} is computed between empirical and theoretical probability distributions.  The empirical failure and survival probabilities are defined as
\begin{equation}
	\left(\hat{q}_{il1}, \hat{q}_{il2}, \hat{q}_{i0}\right)=\left(\frac{n_{il1}}{g_i}, \frac{n_{il2}}{g_i}, \frac{k_i}{g_i}\right),\label{eq5} 
\end{equation}
where, $i=1,2,\dots,I\,;l=1,2,\dots,L.$ The theoretical failure and survival probabilities are given by equation \eqref{eq1}.  The weighted DPD (WDPD) measure with the weights, $w_i=\frac{g_i}{G}$ combining all the I groups is obtained as
\begin{align}
D^w_{\gamma}(\bm{\bm{\Lambda}})=\sum_{i=1}^{I}\frac{g_i}{G}&\left[\left\{
(p_{i0})^{\gamma+1}+\sum_{l=1}^{L}\sum_{r=1}^{2}(p_{ilr})^{\gamma+1}\right\}\right.-\frac{\gamma+1}{\gamma}\bigg\{(\hat{q}_{i0}p_{i0}^{\gamma})+\Bigg.\left.\sum_{l=1}^{L}\sum_{r=1}^{2}\hat{q}_{ilr}(p_{ilr})^{\gamma}\right\}\notag\\
&\qquad+\left.\frac{1}{\gamma}\left\{(\hat{q}_{i0})^{\gamma+1}+\sum_{l=1}^{L}\sum_{r=1}^{2}(\hat{q}_{ilr})^{\gamma+1}\right\}\right] .\label{dpd}
\end{align}  
When, $\gamma\to 0$, $D^w_{\gamma}(\bm{\bm{\Lambda}})$ will converge to Kullback-Leibler (KL) divergence measure.  The weighted minimum DPD estimators (WMDPDE) for estimating $\bm{\bm{\Lambda}}$ is obtained by minimizing the WDPD measure as
\begin{equation}	\hat{\bm{\bm{\Lambda}}}_{\gamma}^w=arg\mathop{min}_{\bm{\bm{\Lambda}}}D^w_{\gamma}(\bm{\bm{\Lambda}}) .\label{eq7}
\end{equation}
The set of estimating equations for obtaining WMDPDE of NOSD test data under Lindley lifetime distribution with a competing risk interval-monitoring set-up is given by
\begin{align}
&\sum_{i=1}^{I}g_i\left[(p_{i0})^{\gamma-1}\left(p_{i0}-\hat{q}_{i0}\right)\frac{\partial(p_{i0})}{\partial\bm{\Lambda}}\right.+\left.\sum_{l=1}^{L}\sum_{r=1}^{2}\left(p_{ilr}\right)^{\gamma-1}\left(p_{ilr}-\hat{q}_{ilr}\right)\frac{\partial(p_{ilr})}{\partial\bm{\Lambda}}\right]=\mathbf{0_4}.\label{eq8}
\end{align}
To study the asymptotic behaviour of WMDPDE, the following theorem inspired by the study of Calvino et al. \cite{cal2021} is presented.
\begin{res}\label{thm1}
	Let $\bm{\Lambda}^0$ be the true value of the parameter $\bm{\Lambda}$.  The asymptotic distribution of WMDPDE of $\bm{\Lambda}$, $\hat{\bm{\Lambda}}_{\gamma}$, is given by
	\begin{equation}
		\sqrt{G}(\hat{\bm{\Lambda}}_{\gamma}-\bm{\Lambda}^0)\xrightarrow[G\to \infty]{\mathscr{L}} N\left(\bm{0_4},Q_{\gamma}^{-1}(\bm{\Lambda}^0)R_{\gamma}(\bm{\Lambda}^0)Q_{\gamma}^{-1}(\bm{\Lambda}^0)\right).\label{9}
	\end{equation}
\end{res}
\begin{proof}
The proof of the result and description of notations are given in the Appendix.
\end{proof}
In the Bayesian context, inspired by Ghosh and Basu \cite{g2016}, we study a robust Bayesian estimation method by deriving pseudo posterior for NOSD under competing risk based on the density power divergence measure in this work.  Let us define the maximize equation based on the WDPD measure for NOSD as
\begin{align}
B^{w}_{\gamma}(\bm{\Lambda})&=\sum_{i=1}^{I}\frac{g_i}{G}\left[\frac{1}{\gamma}\left\{(\hat{q}_{i0}p_{i0}^{\gamma})+\sum_{l=1}^{L}\sum_{r=1}^{2}\hat{q}_{ilr}(p_{ilr})^{\gamma}\right\}\right.\left.-\frac{1}{\gamma+1}\left\{(p_{i0})^{\gamma+1}+\sum_{l=1}^{L}\sum_{r=1}^{2}(p_{ilr})^{\gamma+1}\right\}\right],\label{eq11}
\end{align}
where, WMDPDE with $\gamma>0$ is the maximizer of $B^{w}_{\gamma}(\bm{\Lambda})$.
Therefore, the weighted robust posterior density, a pseudo posterior, can be defined as
\begin{equation}
	\pi^w_{\gamma}(\bm{\Lambda}\vert data)=\frac{\exp\left(B^{w}_{\gamma}(\bm{\Lambda})\right)\pi(\bm{\Lambda})}{\int \exp\left(B^{w}_{\gamma}(\bm{\Lambda})\right)\pi(\bm{\Lambda})\,d\bm{\Lambda}}.\label{eq12}
\end{equation}
Here, $\pi(\bm{\Lambda})$ is the joint prior density, and $\pi^w_{\gamma}(\bm{\Lambda}\vert data)$ is the proper density for $\gamma\geq 0$.
For $\gamma \rightarrow 0$, the robust pseudo posterior will converge to the conventional likelihood-based posterior density.  For any loss-function $L(.,.),$ the Bayes estimator can be obtained as
$$arg  \min_t \int L(\bm{\Lambda}, t) \pi^w_{\gamma}(\bm{\Lambda}\vert data) d \bm{\Lambda}.$$
For the squared error loss function, the Weighted Robust Bayes Estimator (WRBE) can be obtained as
\begin{equation}
	\hat{\bm{\Lambda}}^{w}_{b \gamma}=\int \bm{\Lambda}\pi^w_{\gamma}(\bm{\Lambda}\vert data)\,d\bm{\Lambda}.\label{eq13}
\end{equation}
\subsection{Choice of Priors}
In Bayesian inference, the choice of prior governs the estimation.  In this section, we mention a few such prior choices.  For the model parameters $a_1, a_2, b_1, b_2,$ the interpretation of prior choice is not so meaningful.  Following the idea of Fan et al. \cite{f2009}, the prior information on $p_{ilr}$'s is considered.  We need the empirical estimates of $p_{ilr} $ given in \eqref{eq5} for further development.  Nevertheless, to avoid the zero-frequency situation, we follow the idea of Lee and Morris \cite{le1985} and define
\begin{equation} (\tilde{q}_{i0}, \tilde{q}_{il1}, \tilde{q}_{il2})=\left(\frac{k_i+1}{g_i+2L+1}, \frac{n_{il1}+1}{g_i+2L+1}, \frac{n_{il2}+1}{g_i+2L+1}\right) \label{eq14}.
\end{equation}
\subsubsection{ {Normal Prior based on data}}
Define error $\{e_{ilr} \}$'s as the difference between empirical estimates and the true probabilities as 
\begin{equation}
	\tilde{q}_{ilr}=p_{ilr}+e_{ilr}\;;\quad i=1,2,\dots,I\;;\;l=1,2,\dots,L\;;\;r=1,2,\label{eq15}
\end{equation}
with the assumption that $e_{ilr}$ are independent $N(0,\sigma^2)$ variables.  Therefore, the conditional likelihood function as the prior distribution of $\bm{\Lambda}$ given $\sigma^2$ can be obtained as
\begin{equation}
	L(\bm{\Lambda}\vert \sigma^2)\propto \prod_{i=1}^{I}\prod_{l=1}^{L}\prod_{r=1}^{2}\frac{1}{\sqrt{2\pi\sigma^2}}\exp\left\{-\frac{1}{2\sigma^2}(p_{ilr}-\tilde{q}_{ilr})^2\right\} .
\end{equation}
Using non-informative prior of $\sigma^2$,
$\pi(\sigma^2)\propto \frac{1}{\sigma^2},$
the joint prior density of $\bm{\Lambda}$ is given by
\begin{align}
\pi^{(1)}(\bm{\Lambda})&\propto \int_{0}^{\infty}L(\bm{\Lambda}\vert \bm{\tau}, \bm{s},\sigma^2)\pi(\sigma^2) d\sigma^2\propto \left\{\sum_{i=1}^{I}\sum_{l=1}^{L}\sum_{r=1}^{2}(p_{ilr}-\tilde{q}_{ilr})^2\right\}^{-IL}.\label{nor}
\end{align}
Then, by equation \eqref{eq12}, posterior density would be given as follows.
\begin{equation}
	\pi^{(1)}_{\gamma}(\bm{\Lambda}\vert data)\propto \exp\left(B^w_{\gamma}(\bm{\Lambda})\right)\left\{\sum_{i=1}^{I}\sum_{l=1}^{L}\sum_{r=1}^{2}(p_{ilr}-\tilde{q}_{ilr})^2\right\}^{-IL}. \label{16}
\end{equation}
\subsubsection{{Dirichlet Prior based on data}}
\noindent Beta prior is a natural choice if a parameter can be interpreted as a probability.  Extending this idea similar to Fan et al. \cite{f2009}, a Dirichlet prior is considered for the failure and survival probabilities as
\begin{equation*}	\pi_{i}^{(2)}=\frac{p_{i0}^{\beta_{i0}-1}\prod_{l=1}^{L}\prod_{r=1}^{2}p_{ilr}^{\beta_{ilr}-1}}{\bm{B}(\bm{\beta_i})}\,,
\end{equation*}
where, $\beta_{i0},\beta_{ilr}>0$ for $i=1,\ldots, I;$ $l=1,\ldots, L;$ $r=1,2$ and, $\bm{B}(\bm{\beta_i})=\frac{\Gamma \beta_{i0}\prod_{l=1}^{L}\prod_{r=1}^{2}\Gamma\beta_{ilr}}{\Gamma \left(\beta_{i0}+\sum_{l=1}^{L}\sum_{r=1}^{2}\beta_{ilr}\right)}.$
The hyper-parameters are chosen, equating the expectation of the failure and survival probabilities with their empirical estimates and equating variance as a known constant.  Therefore, we get
\begin{align}
E(p_{i0})&=\frac{\beta_{i0}}{\beta_{i0}+\sum_{l=1}^{L}\sum_{r=1}^{2}\beta_{ilr}}=\tilde{q}_{i0},\;E(p_{ilr})=\frac{\beta_{ilr}}{\beta_{i0}+\sum_{l=1}^{L}\sum_{r=1}^{2}\beta_{ilr}}=\tilde{q}_{ilr}.\label{eq18}\\
Var(p_{i0})&=\frac{\beta_{i0}\left(\sum_{l=1}^{L}\sum_{r=1}^{2}\beta_{ilr}\right)}{\left(\beta_{i0}+\sum_{l=1}^{L}\sum_{r=1}^{2}\beta_{ilr}\right)^2}\times\frac{1}{\left(\beta_{i0}+\sum_{l=1}^{L}\sum_{r=1}^{2}\beta_{ilr}+1\right)}=\sigma^2_{(p)},\label{eq19}
\end{align}
where, $\sigma^2_{(p)}$ is assumed to be a prefixed quantity.  By equations \eqref{eq18} and \eqref{eq19} the estimates of hyper-parameters are
\begin{align*}
\hat{\beta}_{il1}&=\tilde{q}_{il1}\left\{\frac{\tilde{q}_{i0}(1-\tilde{q}_{i0})}{\sigma^2_{(p)}}-1\right\}\;,\quad\hat{\beta}_{il2}=\tilde{q}_{il2}\left\{\frac{\tilde{q}_{i0}(1-\tilde{q}_{i0})}{\sigma^2_{(p)}}-1\right\}\;,\\
\hat{\beta}_{i0}&=\left\{\frac{\tilde{q}_{i0}(1-\tilde{q}_{i0})}{\sigma^2_{(p)}}-1\right\}-\sum_{l=1}^{L}\sum_{r=1}^{2}\hat{\beta}_{ilr}.
\end{align*}
By equation \eqref{eq12}, the joint posterior density would be given as 
\begin{equation}
	\pi^{(2)}_{\gamma}(\bm{\Lambda}\vert data)\propto \exp\left(B^w_{\gamma}(\bm{\Lambda})\right)\prod_{i=1}^{I}\left[p_{i0}^{\hat{\beta}_{i0}-1}\prod_{l=1}^{L}\prod_{r=1}^{2}\left\{p_{ilr}^{\hat{\beta}_{ilr}-1}\right\}\right]\label{20}.
\end{equation}
Under both the prior assumptions, the Bayes estimate cannot be obtained in closed form.  Hence, one can rely on MCMC methods.  Since widely used methods like Gibbs sampler and Metropolis-Hastings (MH) algorithm struggle with high dimensional or highly correlated variables, therefore there has been a growing interest in using the Hamiltonian Monte Carlo (HMC) algorithm for Bayesian estimation recently \cite{thach2020improved,thomas2021learning,zhang2022bayesian,acharyya2024monotone}.  The HMC steps are given in the algorithm \ref{alg}.
\begin{algorithm}[htb!]
\caption{{{Hamiltonian Monte Carlo}}}\label{alg}
\begin{itemize}
\item Define the diagonal matrix $\bm{M}$, step size $\epsilon$, leapfrog step $L$ and sample size $N$.
\item Initialize the position state $\bm{\Lambda}^{(0)}.$
\item[] For $t=1,2,\dots,N$
\item Sample $\bm{\phi}^{(t)}\sim N(\bm{0},\bm{M}).$
\item Run leapfrog starting at $(\bm{\Lambda}^{(t)},\bm{\phi}^{(t)})$ for $L$ step with step size $\epsilon$ to produce proposed state $(\bm{\Lambda}^{*},\bm{\phi}^{*})$.
\item[] Let $\bm{\phi}^{(t,0)}=\bm{\phi}^{(t)}$ and  $\bm{\Lambda}^{(t-1,0)}=\bm{\Lambda}^{(t-1)}$, then for $t^{'}=1,2,\dots,N$
\item $\bm{\phi}_{\epsilon/2}=\bm{\phi}^{(t,t^{'}-1)}+\left.\frac{\epsilon}{2}\frac{\partial \log\pi_{\alpha}(\bm{\Lambda}\vert t)}{\partial\bm{\Lambda}}\right\vert_{\bm{\Lambda}=\bm{\Lambda}^{(t-1,t^{'}-1)}}$
\item $\bm{\Lambda}^{t-1,t^{'}}=\bm{\Lambda}^{t-1,t^{'}-1}+\epsilon\, \bm{M}^{-1}\bm{\phi}_{\epsilon/2}$
\item $\bm{\phi}^{t,t^{'}}=\bm{\phi}_{\epsilon/2}+\left.\frac{\epsilon}{2}\frac{\partial \log\pi_{\alpha}(\bm{\Lambda}\vert t)}{\partial\bm{\Lambda}}\right\vert_{\bm{\Lambda}=\bm{\Lambda}^{(t-1,t^{'})}}$
\item[] Hence, $\bm{\Lambda}=\bm{\Lambda}^{t-1,L}$ and $\phi^{*}=\phi^{t,L}$.
\item Compute acceptance probability\\ $acc=min\Big\{1,\exp\big(U(\bm{\Lambda}^{(t-1)})-U(\bm{\Lambda}^{*})+K(\bm{\phi}^{(t)})\big)-K(\bm{\phi}^{*})\big)\Big\}.$
\item Generate a random number $u\sim U(0,1)$ and set\\
$\bm{\theta}^{(t)}=\begin{cases}
\bm{\theta}^{*}&; \;u\leq acc.\\
\bm{\theta}^{(t-1)}&;\; \text{otherwise}.
\end{cases} $
\item Stop when $t=N$.
\end{itemize}
\end{algorithm}

\section{Testing of hypothesis based on Robust Bayes Factor}\label{sec4}
To validate if the available data supports a particular hypothesis is often a study of interest.  When outliers are present in the data set, robust hypothesis testing is prudent.  Another essential contribution of this study is to develop the testing of the hypothesis based on the robust Bayes factor in the context of NOSD under competing risk. 
 Inspired by the procedure followed by Ghosh et al. \cite{ghosh2006introduction}, let the parameter space of 
$\bm{\Lambda}$ be denoted by $\bm{\Theta}$.  Based on the values of a vector-valued function $f_n:\mathbb{R}^{4}\xrightarrow[]{}\mathbb{R}^{d}, \quad d\leq 4$ the null and the alternative hypotheses are given as
\begin{align*}
	\bm{H}_0 :& \bm{\Lambda}\in\bm{\Theta}_0=\{\bm{\Lambda}\in\bm{\Theta} : f_n(\bm{\Lambda})=\bm{0}_d\}\quad \text{against}\\
 \bm{H}_1 :& \bm{\Lambda}\in\bm{\Theta}_1=\{\bm{\Lambda}\notin\bm{\Theta}_0\}.
\end{align*}
Further, let $\rho_0$ and $1-\rho_0$ be the prior probabilities under $\bm{\Theta}_0$ and $\bm{\Theta}_1$ respectively.  Let $\pi_0(\bm{\Lambda})$ and $\pi_1(\bm{\Lambda})$ be the prior density of $\bm{\Lambda}$ under $\bm{\Theta}_0$ and $\bm{\Theta}_1$ respectively, such that, $\int_{\bm{\Theta}_i}\pi_i(\bm{\Lambda})d\bm{\Lambda}=1$.  Then, prior can be written as
\begin{equation*}
	\pi(\bm{\Lambda})=\rho_0\pi_0(\bm{\Lambda}) I\{\bm{\Lambda}\in\bm{\Theta}_0\}+(1-\rho_0)\pi_1(\bm{\Lambda}) I\{\bm{\Lambda}\in\bm{\Theta}_1\}.
\end{equation*}
The marginal density under the prior $\pi$ can be expressed as
\begin{align*}
M_{\gamma}(\pi)&=\rho_0\int_{\bm{\Theta}_0}\exp(B^w_{\gamma}(\bm{\Lambda}))\pi_0(\bm{\Lambda})d\bm{\Lambda}+(1-\rho_0)\int_{\bm{\Theta}_1}\exp(B^w_{\gamma}(\bm{\Lambda}))\pi_1(\bm{\Lambda})\,d\bm{\Lambda}.
\end{align*}
Hence, posterior density is obtained as
\begin{align*}
	\pi_{\gamma}(\bm{\Lambda}\vert data)&=\frac{\exp(B^w_{\gamma}(\bm{\Lambda}))\pi(\bm{\Lambda})}{M_{\gamma}(\pi)}\\
	&=
	\begin{cases}
		\frac{\rho_0 \exp(B^w_{\gamma}(\bm{\Lambda}))\pi_0(\bm{\Lambda})}{M_{\gamma}(\pi)} &if\; \bm{\Lambda}\in\bm{\Theta}_0\\
		\frac{(1-\rho_0) \exp(B^w_{\gamma}(\bm{\Lambda}))\pi_1(\bm{\Lambda})}{M_{\gamma}(\pi)} &if\; \bm{\Lambda}\in\bm{\Theta}_1.\\
	\end{cases}
\end{align*}
Therefore, we derive the posterior probabilities under $\bm{\Theta}_0$ and $\bm{\Theta}_1$ as
\begin{align*}
	P_{\pi_{\gamma}}(\bm{\Lambda}\in\bm{\Theta}_0\vert data)&=\frac{\rho_0}{M_{\gamma}(\pi)}\int_{\bm{\Theta}_0}\exp(B^w_{\gamma}(\bm{\Lambda}))\pi_0(\bm{\Lambda})\,d\bm{\Lambda},\\
	P_{\pi_{\gamma}}(\bm{\Lambda}\in\bm{\Theta}_1\vert data)&=\frac{(1-\rho_0)}{M_{\gamma}(\pi)}\int_{\bm{\Theta}_1}\exp(B^w_{\gamma}(\bm{\Lambda}))\pi_1(\bm{\Lambda})\,d\bm{\Lambda}.
\end{align*}
The posterior odds ratio of $H_0$ relative to $H_1$ is given by
\begin{equation}
	\frac{ P_{\pi_{\gamma}}(\bm{\Lambda}\in\bm{\Theta}_0\vert data)}{ P_{\pi_{\gamma}}(\bm{\Lambda}\in\bm{\Theta}_1\vert data)}=\left(\frac{\rho_0}{1-\rho_0}\right)BF_{01},
\end{equation}
where, $BF_{01}$ is the Bayes factor given as
\begin{equation}
	BF_{01}=\frac{\int_{\bm{\Theta}_0}\exp(B^w_{\gamma}(\bm{\Lambda}))\pi_0(\bm{\Lambda})\,d\bm{\Lambda}}{\int_{\bm{\Theta}_1}\exp(B^w_{\gamma}(\bm{\Lambda}))\pi_1(\bm{\Lambda})\,d\bm{\Lambda}}.\label{bf}
\end{equation}
The smaller value of $BF_{01}$ indicates the stronger evidence against $H_0$.

\section{Property of Robustness}\label{sec5}
In previous sections, we have derived the estimators of the model parameters using the robust estimation method.  In this section, the robustness of those estimators will be studied through the influence function (IF) \cite{law1986robust,hm2011}.  Suppose $S(M)$ denotes the functional of any estimator of any true distribution M, then the IF is represented as
\begin{equation*}
	IF(t;S,M)=\lim_{\epsilon \to 0} \frac{S(M_{\epsilon}) - S(M) }{\epsilon} =\left. \frac{\partial S(M_{\epsilon}) }{\partial \epsilon} \right\vert_{\epsilon =0},
\end{equation*}
where, $M_{\epsilon}=(1-\epsilon)M+\epsilon\Delta_{t}$, $\epsilon (0 < \epsilon < 1)$ be the proportion of contamination and $\Delta_t$ be the degenerate
distribution at the point t.  Let $F_{\bm{\Lambda}}$ be the true distribution from which data is generated.  If $S_{\gamma}(F_{\bm{\Lambda}})$ be the functional of WMDPDE $\hat{\bm{\Lambda}}_{\gamma}$, then the influence function of $\hat{\bm{\Lambda}}_{\gamma}$ based on all I groups is given as follows.
\begin{align*}
IF(\bm{t;S_{\gamma},F_{\Lambda}}){=}\frac{Q^{-1}_{\gamma}(\bm{\Lambda})}{G}\sum_{i=1}^{I}g_i\left[\big\{\delta_{I_{i0}}(t)-p_{i0}\big\}p^{\gamma-1}_{i0}\frac{\partial(p_{i0})}{\partial\bm{\Lambda}}\right.\left.+\sum_{l=1}^{L}\sum_{m=1}^{2}\big\{\delta_{I_{ilr}}(t)-p_{ilr}\big\}p^{\gamma-1}_{ilr}\frac{\partial(p_{ilr})}{\partial\bm{\Lambda}}\right],
\end{align*}
 where,
 $\delta_{I_A}(\bm{t}) = 
\begin{cases}
	1 \quad \text{if} \  \bm{t} \in I_A\\
	0 \quad \text{otherwise}\\
\end{cases}
$.

\subsection{Influence Function of WRBE}
The robustness property corresponding to robust Bayes estimator $\hat{\bm{\Lambda}}^w_{b \gamma}$ using IF \cite{g2016} is studied through Bayes functional, which is given as follows concerning squared error loss function.
\begin{equation*}
T^{(\gamma)}_G(F_{\bm{\Lambda}})=\frac{\int \bm{\Lambda}\exp\left\{ B^w_{\gamma}(\bm{\Lambda};F_{\bm{\Lambda}})\right\}\pi(\bm{\Lambda})d\bm{\Lambda}}{\int\exp\left\{B^w_{\gamma}(\bm{\Lambda};F_{\bm{\Lambda}})\right\}\pi(\bm{\Lambda})d\bm{\Lambda}},
\end{equation*}
where, $B^w_{\gamma}(\bm{\Lambda};F_{\bm{\Lambda}})$ is defined as
\begin{flalign*}
B^w_{\gamma}(\bm{\Lambda};F_{\bm{\Lambda}})&=\frac{1}{\gamma}\sum_{i=1}^{I}\left[\left\{\left(\int_{I_{i0}}dF_{\bm{\Lambda}}\right)p^{\gamma}_{i0}+\sum_{l=1}^{L}\sum_{r=1}^{2}\right.\right.\left.\left.\left(\int_{I_{ilr}}dF_{\bm{\Lambda}}\right)p^{\gamma}_{ilr}\right\}\right.\left.-\frac{1}{\gamma+1}\left\{p^{\gamma+1}_{i0}+\sum_{l=1}^{L}\sum_{r=1}^{2}p^{\gamma+1}_{ilr}\right\}\right].
&&
\end{flalign*}
Based on the above discussion, the following result provides the influence function of the WRBE.
\begin{res}\label{res4}
The influence function of $\hat{\bm{\Lambda}}^w_{ b\gamma}$,  based on all I groups, is given by
\begin{equation*}
IF(\bm{\bm{t};T_G^{(\gamma)},F_{\bm{\Lambda}}})=Cov_{(p)}\left(\bm{\Lambda},X_{\gamma}(\bm{\Lambda};\bm{t},f_{\bm{\Lambda}})\right).
\end{equation*}
 where,
\begin{align*}
X_{\gamma}(\bm{\Lambda};t,f_{\bm{\Lambda}})&=\frac{1}{\gamma}\sum_{i=1}^{I}\frac{g_i}{G}\left[\bigg\{\big(\delta_{I_{i0}}(t)-p_{i0}\big)p_{i0}^{\gamma}\bigg\}\right.\left.+\left\{\sum_{l=1}^{L}\sum_{r=1}^{2}\big(\delta_{I_{ilr}}(t)-p_{ilr}\big)p^{\gamma}_{ilr}\right\}\right].
\end{align*}
\end{res}
\begin{proof}
Given in the Appendix.
\end{proof}

\subsection{Influence Function of Bayes Factor}
In this section, the robustness property of the Bayes factor is examined by deriving its influence factor when the null hypothesis is true.  Let $F_{\Lambda_0}$ be the true distribution under the null hypothesis $H_0: \bm{\Lambda}\in\bm{\Theta}_0$  and therefore the functional related to the Bayes factor can be defined as
\begin{equation*}
	T^{(\gamma)}_{\bm{\Theta}}(F_{\bm{\Lambda}_0})=\frac{\int_{\bm{\Theta}_0}\exp\big\{B^w_{\gamma}(\bm{\Lambda}\in\bm{\Theta}_0;F_{\bm{\Lambda}_0})\big\}\pi_0(\bm{\Lambda})\,d\bm{\Lambda}}{\int_{\bm{\Theta}_1}\exp\big\{B^w_{\gamma}(\bm{\Lambda}\in\bm{\Theta}_1;F_{\bm{\Lambda}_0})\big\}\pi_1(\bm{\Lambda})\,d\bm{\Lambda}}.
\end{equation*}
Here, $B^w_{\gamma}(\bm{\Lambda}\in\bm{\Theta}_j;F_{\bm{\Lambda}_0})$ is expressed as
\begin{flalign*}
B^w_{\gamma}(\bm{\Lambda}\in\bm{\Theta}_j;F_{\bm{\Lambda}_0})&=\frac{1}{\gamma}\sum_{i=1}^{I}\left[\left\{\left(\int_{I_{i0}}dF_{\bm{\Lambda}_0}\right)p^{\gamma}_{i0}(\bm{\Lambda}\in\bm{\Theta}_j)+\sum_{l=1}^{L}\sum_{r=1}^{2}\left(\int_{I_{ilr}}dF_{\bm{\Lambda}_0}\right)p^{\gamma}_{ilr}(\bm{\Lambda}\in\bm{\Theta}_j)\right\}\right.\\
&\qquad-\frac{1}{\gamma+1}\left.\left\{p^{\gamma+1}_{i0}(\bm{\Lambda}\in\bm{\Theta}_j)+\sum_{l=1}^{L}\sum_{r=1}^{2}p^{\gamma+1}_{ilr}(\bm{\Lambda}\in\bm{\Theta}_j)\right\}\right]\;;\;j=0,1.
&&
\end{flalign*}
Let the contamination in the true distribution $F_{\Lambda_0}$ under $H_0: \bm{\Lambda}\in\bm{\Theta}_0$ be given as $M_{\epsilon}=(1-\epsilon)F_{\bm{\Lambda}_0}+\epsilon\Delta_t$.  Then IF with respect to the contamination $M_{\epsilon}$ is obtained as
\begin{equation*}
IF(t;T^{(\gamma)}_{\bm{\Theta}},F_{\bm{\Lambda}_0})=\left.\frac{\partial(T^{(\gamma)}_{\bm{\Theta}}(M_{\epsilon}))}{\partial\epsilon}\right\vert_{\epsilon\to 0^{+}}.
\end{equation*}
The following results provide the explicit expression of the IF under this given setup.

\begin{res}\label{res5}
	Based on all I groups, the influence function of Bayes factor BF$_{01}$ is given as follows.
	\begin{equation*}
		IF(t;T^{(\gamma)}_{\bm{\Theta}},F_{\bm{\Lambda}_0})=Y_{\gamma}(\bm{\Theta})\bigg\{E\Big[X_{\gamma}(\bm{\Lambda}\in\bm{\Theta}_0)\Big]-E\Big[X_{\gamma}(\bm{\Lambda}\in\bm{\Theta}_1)\Big]\bigg\}.
	\end{equation*}
 where,
 \begin{flalign*}
     Y_{\gamma}(\bm{\Theta})=\frac{\int_{\bm{\Theta}_0}\exp\big\{B^w_{\gamma}(\bm{\Lambda}\in\bm{\Theta}_0)\big\}\pi_0(\bm{\Lambda})\,d\bm{\Lambda}}{\int_{\bm{\Theta}_1}\exp\big\{B^w_{\gamma}(\bm{\Lambda}\in\bm{\Theta}_1)\big\}\pi_1(\bm{\Lambda})\,d\bm{\Lambda}}.
     &&
 \end{flalign*}
\end{res}
\begin{proof}
Given in the Appendix.
\end{proof}

\section{Numerical Study}\label{sec6}
\noindent In this section, the performance of the theoretical results developed in previous sections has been assessed numerically through simulation experiments and real data analysis.
\subsection{Simulation Analysis}
\begin{figure}[htb!]
\centering
\subfloat[\textrm{\scriptsize{Group 1 (Absolute bias)}}]{\includegraphics[height=3cm,width =0.31\textwidth]{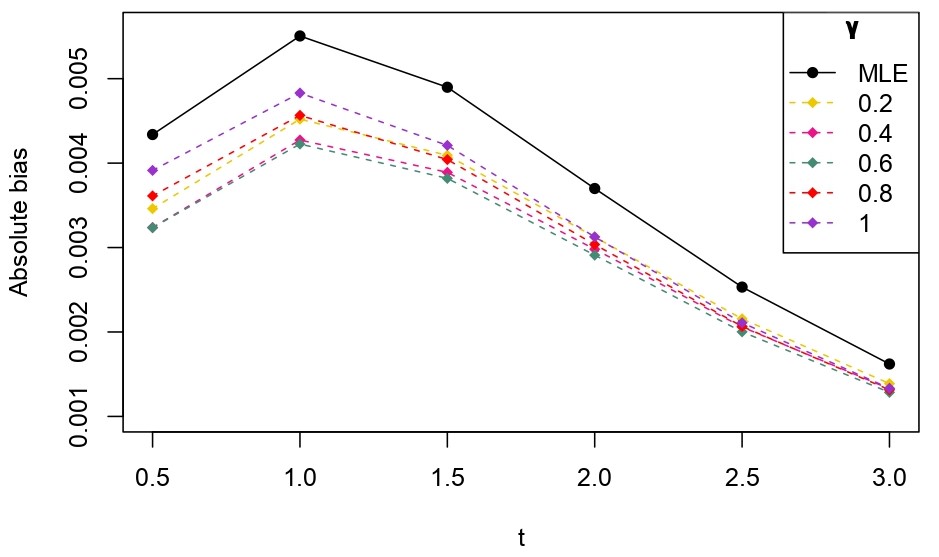}} 
\subfloat[\textrm{\scriptsize{Group 2 (Absolute bias)}}]{\includegraphics[height=3cm,width =0.31\textwidth]{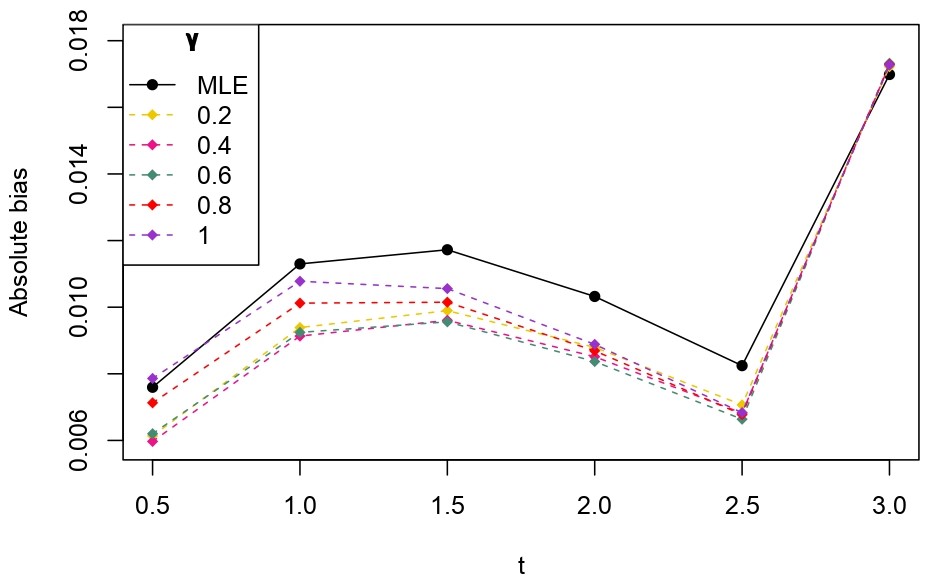}}
\subfloat[\textrm{\scriptsize{Group 3 (Absolute bias)}}]{\includegraphics[height=3cm,width =0.31\textwidth]{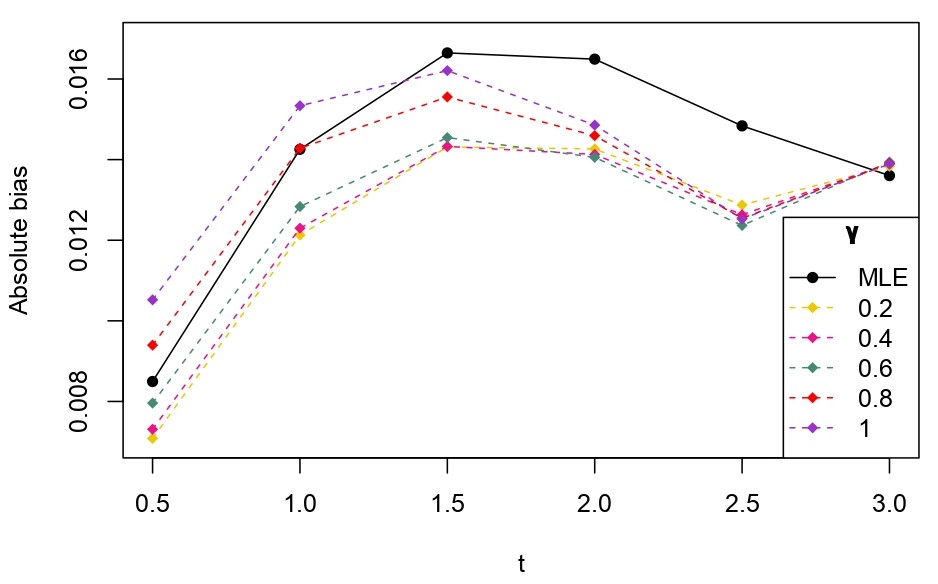}}\\
\subfloat[\textrm{\scriptsize{Group 1 (RMSE)}}]{\includegraphics[height=3cm,width =0.31\textwidth]{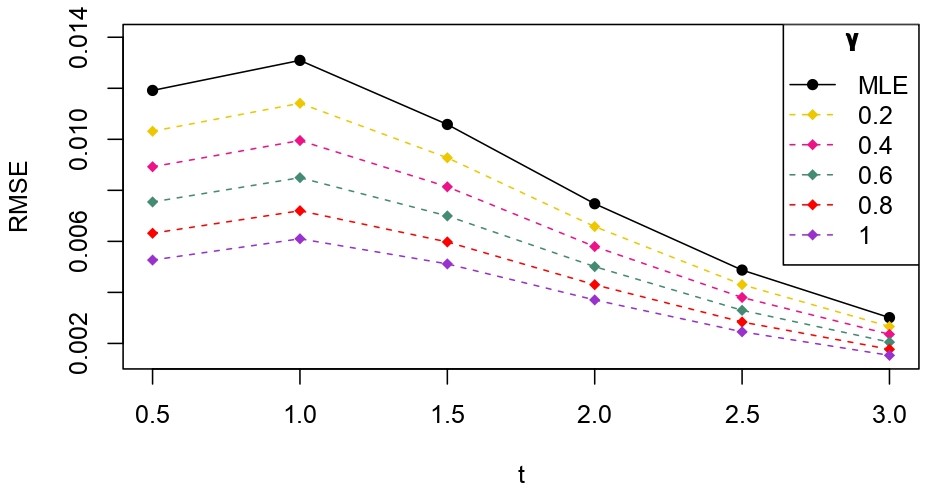}}
\subfloat[\textrm{\scriptsize{Group 2 (RMSE)}}]{\includegraphics[height=3cm,width =0.31\textwidth]{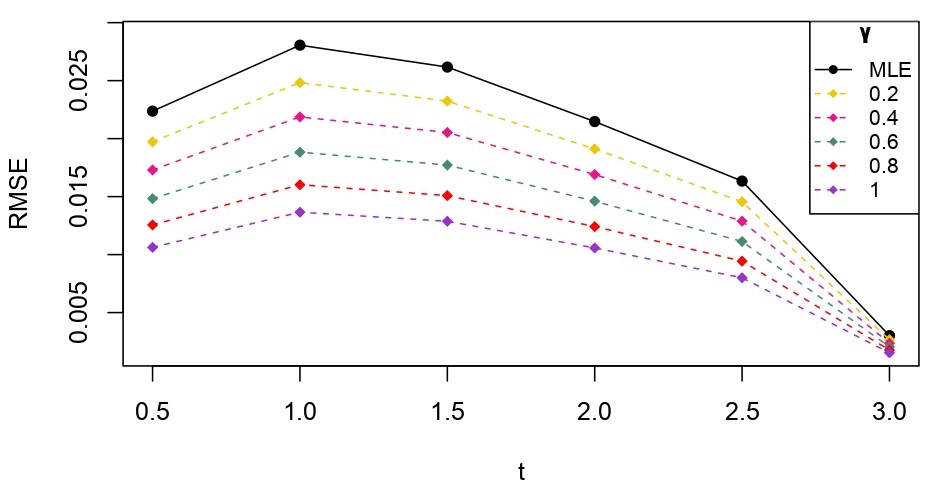}} 
\subfloat[\textrm{\scriptsize{Group 3 (RMSE)}}]{\includegraphics[height=3cm,width =0.31\textwidth]{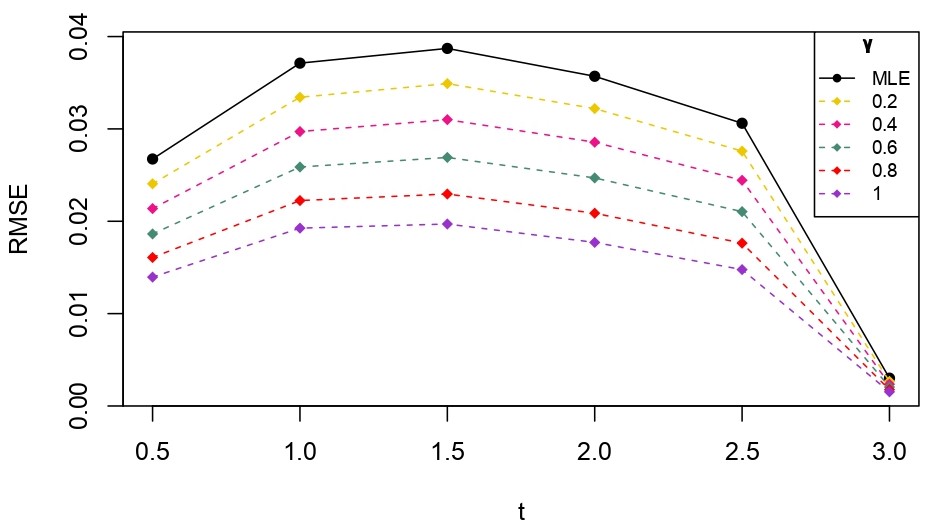}}
\caption{Absolute bias and RMSE for reliability estimates of MLE and WMDPDE under contamination.}
\label{fig1}
\end{figure}

\begin{figure}[htb!]
\centering
\subfloat[\textrm{\scriptsize{Group 1 (Absolute bias)}}]{\includegraphics[height=3cm,width =0.31\textwidth]{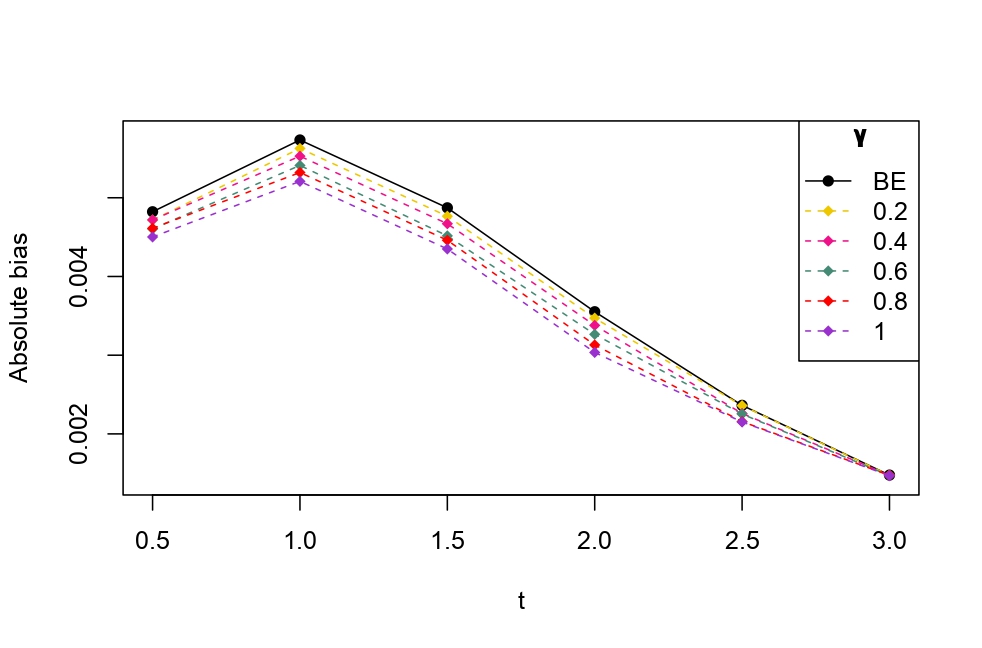}} 
 \subfloat[\textrm{\scriptsize{Group 2 (Absolute bias)}}]{\includegraphics[height=3cm,width =0.31\textwidth]{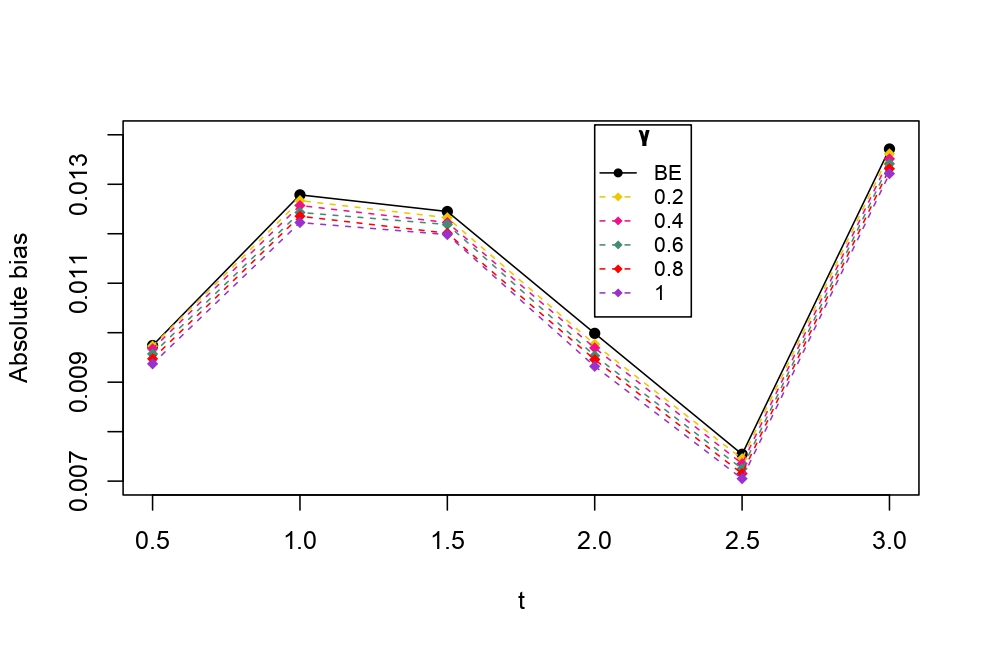}}
 \subfloat[\textrm{\scriptsize{Group 3 (Absolute bias)}}]{\includegraphics[height=3cm,width =0.31\textwidth]{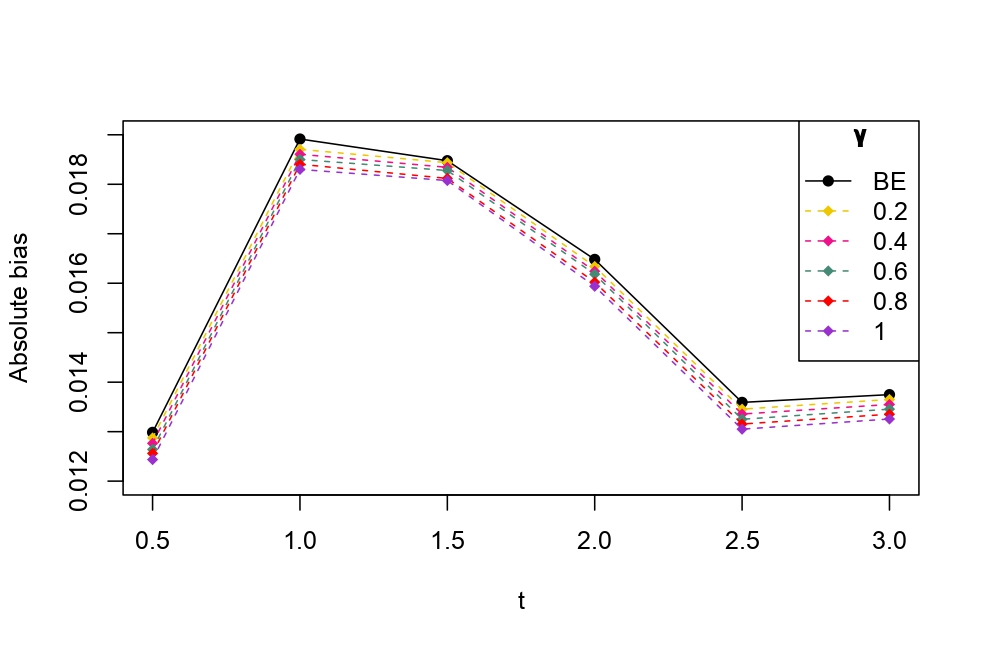}}\\
\subfloat[\textrm{\scriptsize{Group 1 (RMSE)}}]{\includegraphics[height=3cm,width =0.31\textwidth]{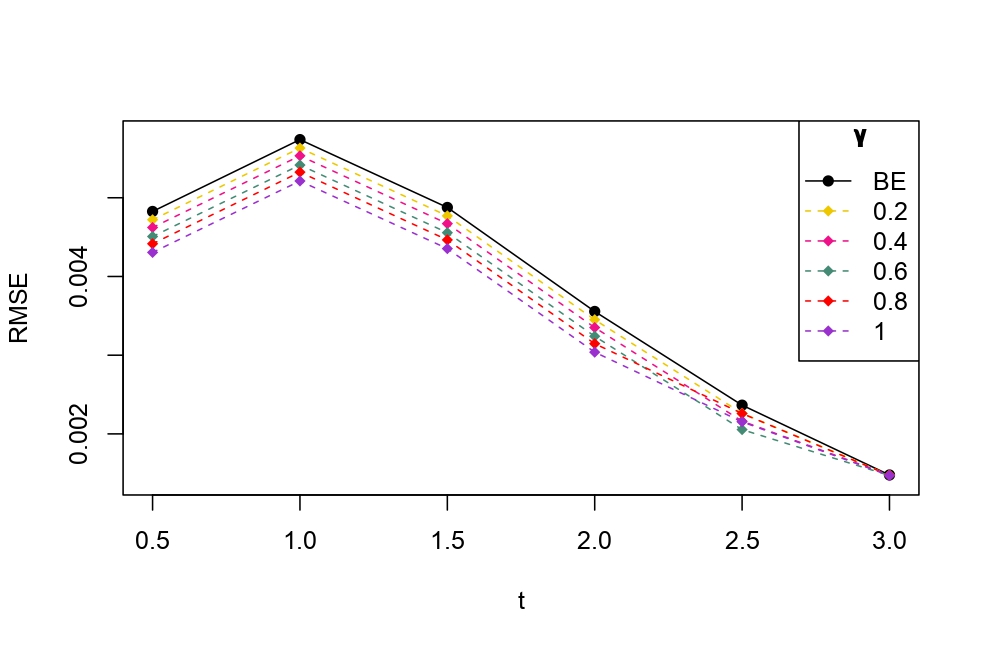}}
\subfloat[\textrm{\scriptsize{Group 2 (RMSE)}}]{\includegraphics[height=3cm,width =0.31\textwidth]{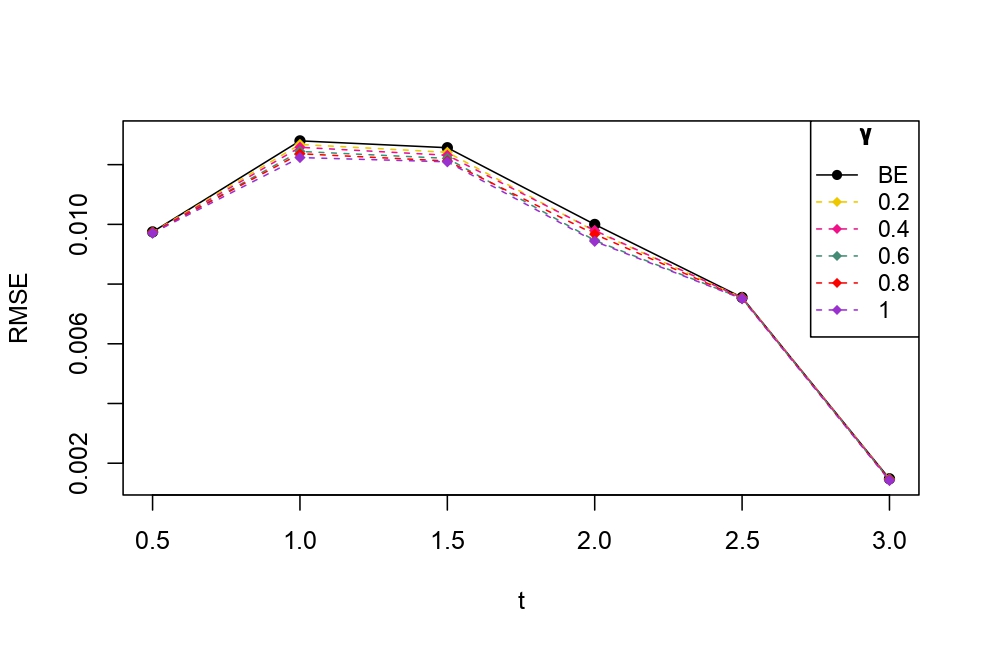}}
\subfloat[\textrm{\scriptsize{Group 3 (RMSE)}}]{\includegraphics[height=3cm,width =0.31\textwidth]{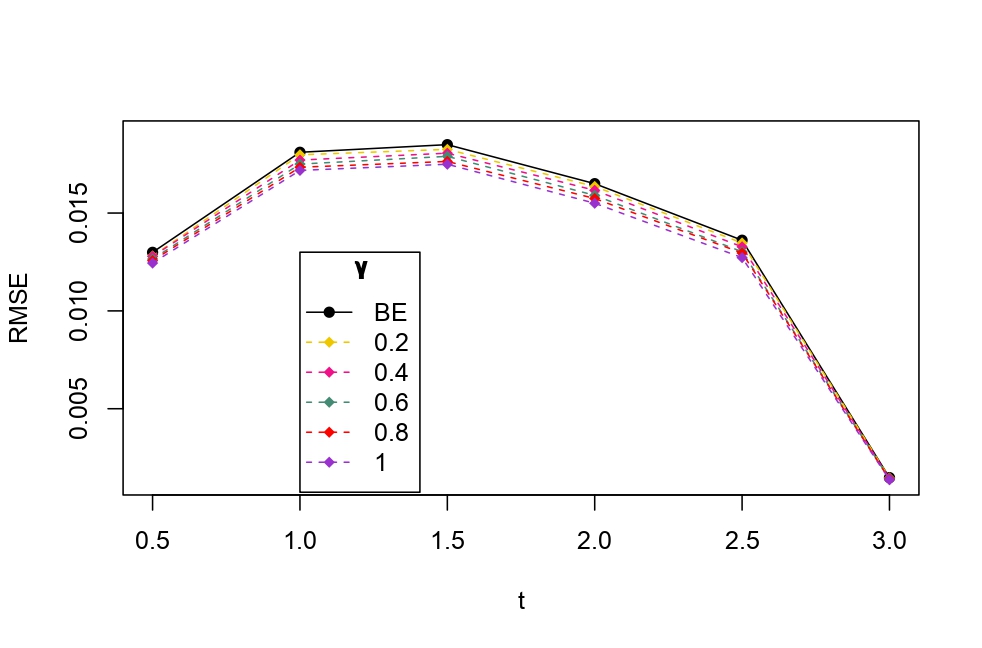}}
\caption{Absolute bias and RMSE for reliability estimates of BE and WRBE (Normal prior) under contamination.}
	\label{fig2}
\end{figure}

\begin{figure}[htb!]
\centering
\subfloat[\textrm{\scriptsize{Group 1 (Absolute bias)}}]{\includegraphics[height=3cm,width =0.31\textwidth]{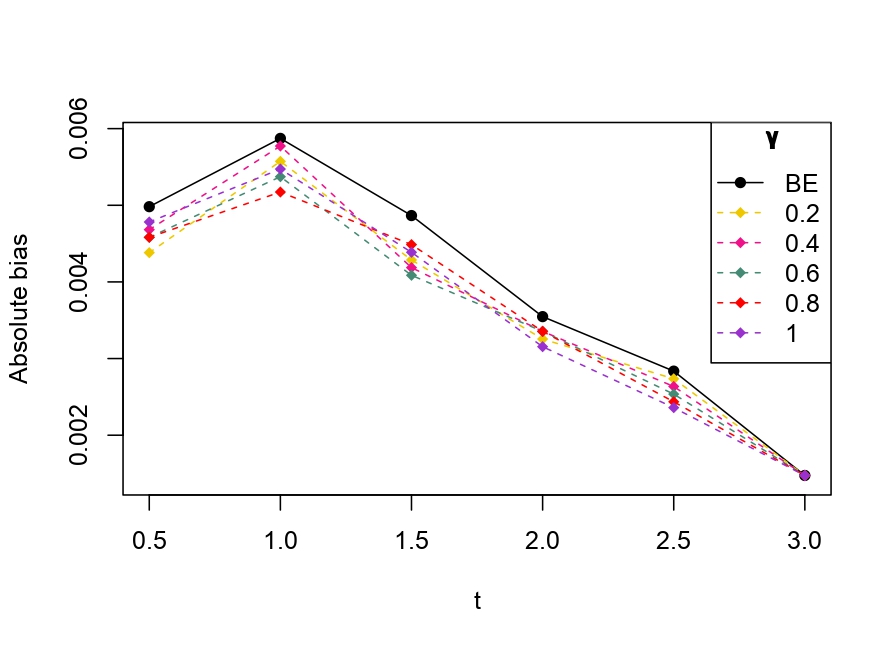}} 
 \subfloat[\textrm{\scriptsize{Group 2 (Absolute bias)}}]{\includegraphics[height=3cm,width =0.31\textwidth]{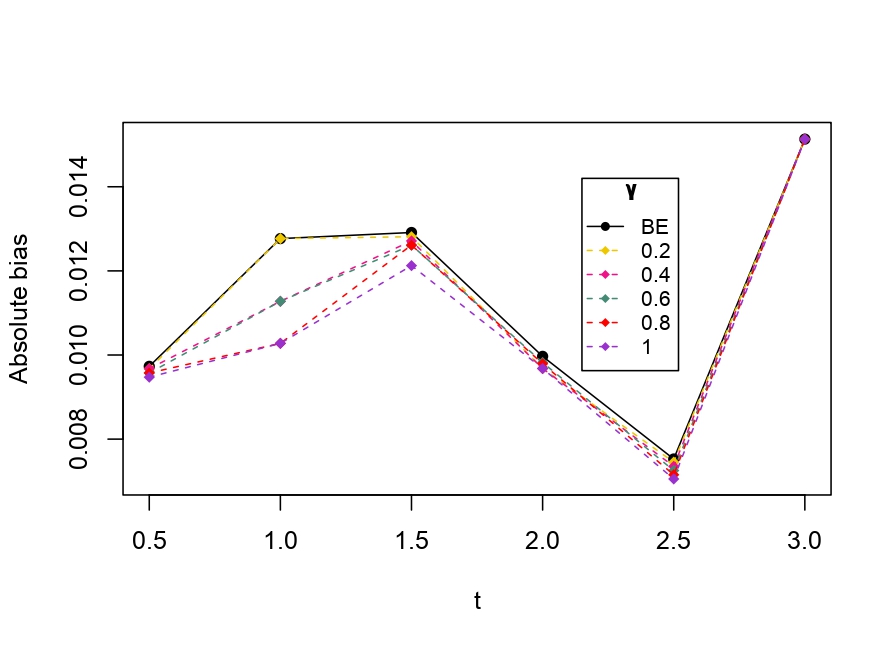}}
  \subfloat[\textrm{\scriptsize{Group 3 (Absolute bias)}}]{\includegraphics[height=3cm,width =0.31\textwidth]{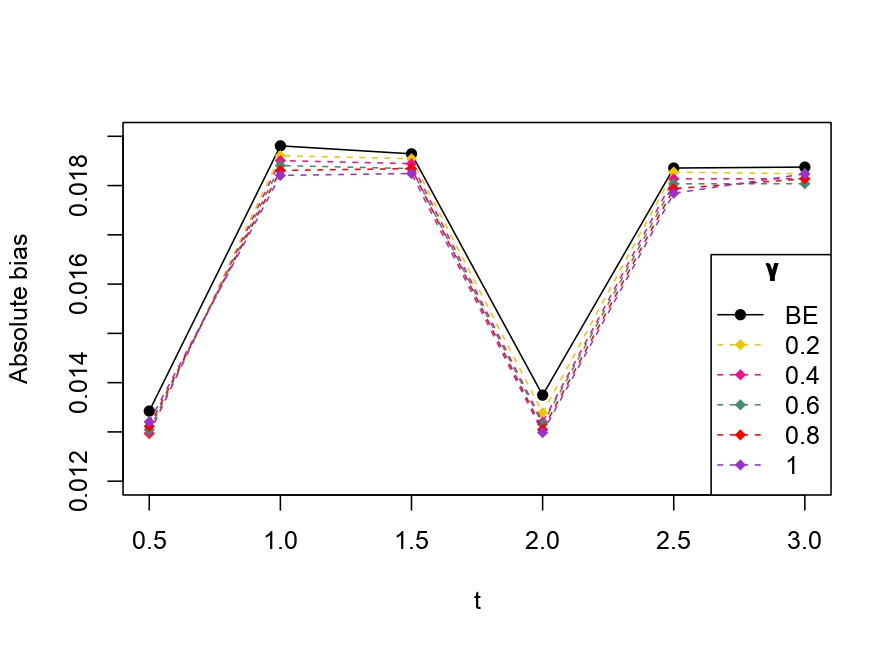}}\\
\subfloat[\textrm{\scriptsize{Group 1 (RMSE)}}]{\includegraphics[height=3cm,width =0.31\textwidth]{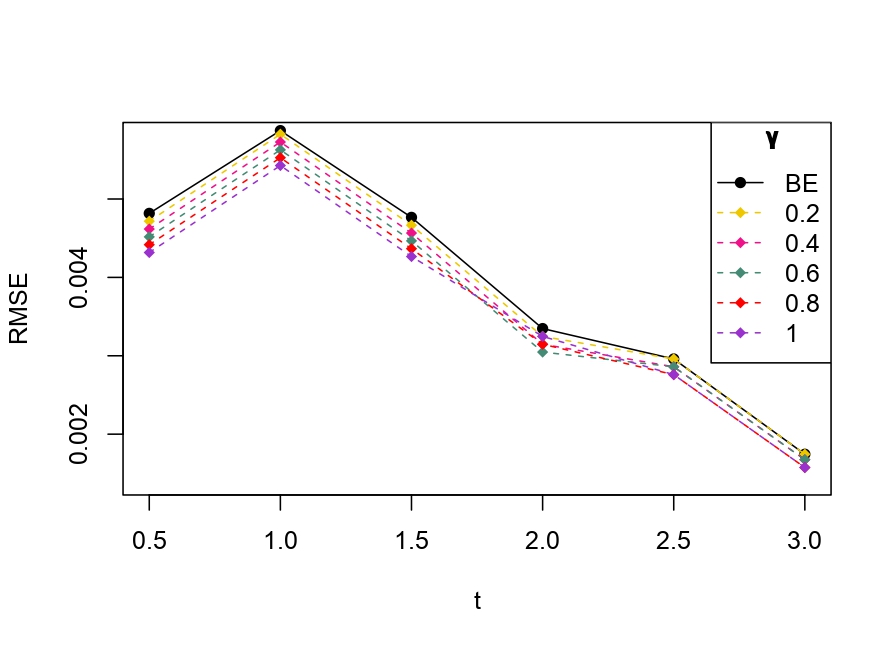}}
\subfloat[\textrm{\scriptsize{Group 2 (RMSE)}}]{\includegraphics[height=3cm,width =0.31\textwidth]{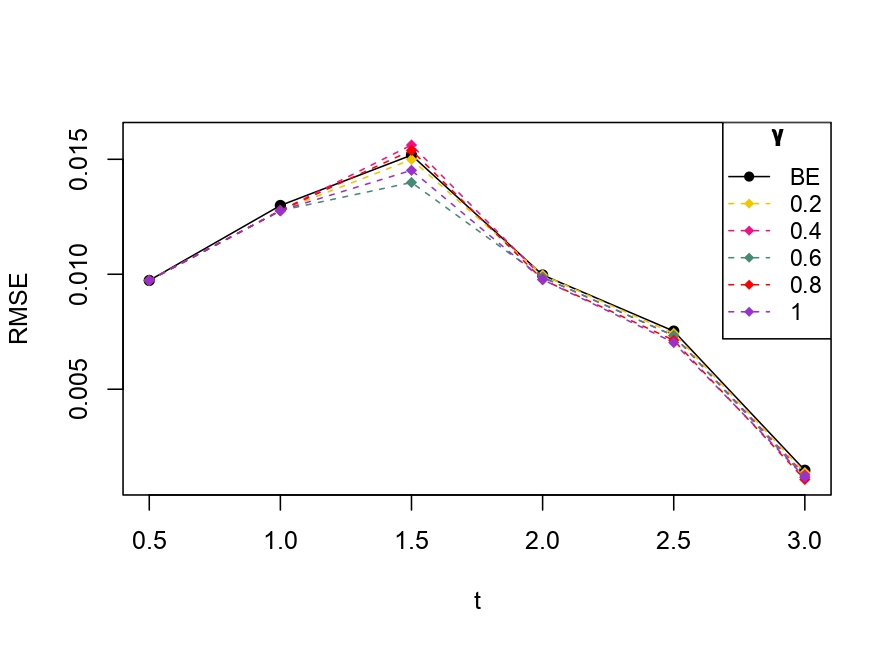}} 
\subfloat[\textrm{\scriptsize{Group 3 (RMSE)}}]{\includegraphics[height=3cm,width =0.31\textwidth]{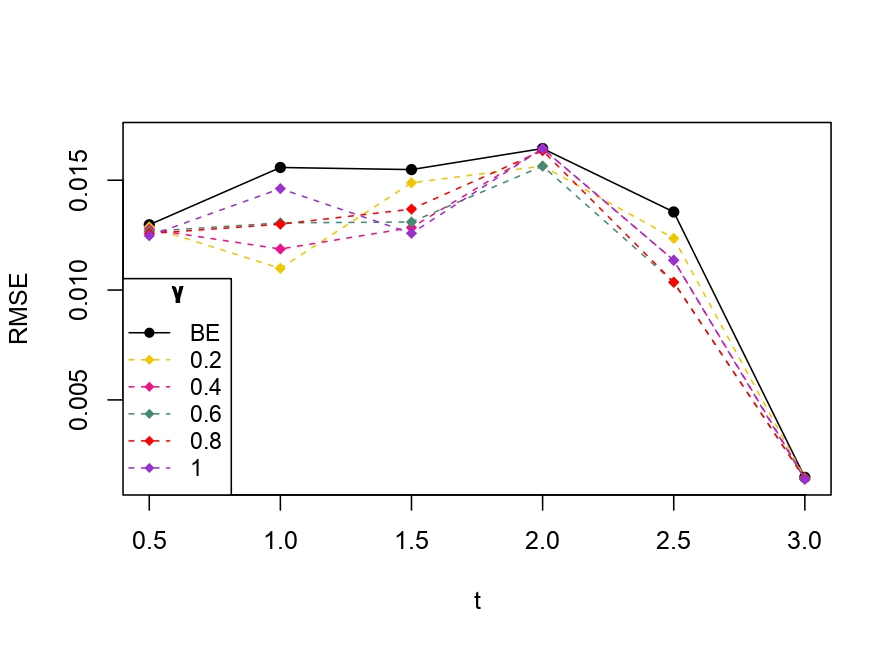}}
\caption{Absolute bias and RMSE for reliability estimates of BE and WRBE (Dirichlet prior) under contamination.}
\label{fig3}
\end{figure}

For simulation purposes, 75 non-destructive one-shot devices (NOSD) following Lindley Lifetime distribution are put to the ALT experiment under two competing causes of failures.  The layout for two simulation experiments (Sim. 1 and Sim. 2) are given in Table \ref{tab2}.  The true values of model parameters to generate the pure data are taken as $\bm{\Lambda}_1{=}(-0.20, -0.06, 0.30, -0.17)^{'}$ and $\bm{\Lambda}_2{=}(-0.11, 0.11, -0.68, 0.09)^{'}$ for two simulation schemes respectively.  Here, the contamination scheme is to generate data from any contaminated version of the true distribution as $\tilde{\bm{\Lambda}}_1{=}(a_1{-}0.01, b_1{-}0.01, a_2{+}0.02, b_2{+}0.02)^{'}$ and $\tilde{\bm{\Lambda}}_2{=}(a_1{+}0.009, b_1{+}0.02, a2{-}0.02, b2{-}0.009)^{'}$.
\begin{table}[htb!]
\tbl{Model layout for the simulation study.}
{\scalebox{1}{
{\begin{tabular}{lccccc} \toprule
{Groups}&{Devices} & \multicolumn{2}{c}{{Stress Levels}}& \multicolumn{2}{c}{{Inspection Times}} \\ \cmidrule{3-6}
& & {Sim. 1}& {Sim. 2} & {Sim. 1}& {Sim. 2}\\ \midrule
1&20 &1.5 &2.0&(0.1,0.7,1.6)&(0.1,0.5,1.0)  \\
2&25 &3.5&4.0 &(0.3,1.0,2.7)&(0.2,0.7,2.0) \\
3&30 &5.5& 6.0&(0.3,1.0,3.0) &(0.3,0.6,1.0) \\ \bottomrule
\end{tabular}}}}
\label{tab2}
\end{table}
Robustness behaviour can be observed through the study of the biases of the estimators.  Hence, the Bias of MLE and WMDPDE is obtained through Monte Carlo simulation based on 1000 generations where MLE and WMDPDE are obtained using the coordinate descent algorithm \cite{baghel2024analysis,baghel2024robust}.  The outcomes are reported in Table \ref{tab3} in pure and contaminated cases.  For Bayesian estimation, Bayes estimate (BE) and robust Bayes estimate (RBE) are obtained by using Hamiltonian Monte Carlo (HMC) given in the algorithm \ref{alg}.  For the smooth running of the HMC algorithm, we consider step size $\epsilon=(0.001,0.05)$, number of steps $L=(2,5)$ and define $v=(0.05,0.05,0.05,0.05); (0.1,0.08,0.006,0.005)$ for two schemes, respectively.  $\bm{M}$ is taken as diagonal matrix whose diagonal elements are $1/v$.  Two chains of $N=1200$ values are generated through HMC, and the first $N^{'}=200$ values are discarded as burn-in period.    The bias for BE and RBE under normal and Dirichlet prior are reported in table \ref{tab4}, where for Dirichlet prior $\sigma^2_{(p)}=0.06$ is taken.
\begin{table}[htb!]
\tbl{Bias of MLE and WMDPDE under simulation.}
{\scalebox{1}{
{\begin{tabular}{lcccccccc} \toprule
&\multicolumn{4}{c}{{Pure Data}} & \multicolumn{4}{c}{{Contamination}} \\\cmidrule(lr){2-5}\cmidrule(lr){6-9}
& $\hat{a}_1$ & $\hat{b}_1$ & $\hat{a}_2$ & $\hat{b}_2$&  $\tilde{a}_1$ & $\tilde{b}_1$ & $\tilde{a}_2$ & $\tilde{b}_2$\\ \midrule
\multicolumn{9}{l}{$\bm{\Lambda}=\bm{\Lambda}_1$}\\
\midrule
{MLE} & -0.003111 & -0.001370   & 0.005675 & 0.000684& 0.030166 & -0.019073  &-0.022712   &0.014589\\\midrule
\multicolumn{9}{l}{{WMDPDE}}  \\
\midrule
$\mathbf{\gamma=0.2}$ & 0.004071& -0.004281& -0.016899& -0.000776& 0.020468& -0.013994  &-0.012792   &0.008672\\
$\mathbf{\gamma=0.4}$&  0.007240& -0.002899& -0.014785 &-0.000799&0.024679& -0.010748&-0.012095  & 0.008011 \\
$\mathbf{\gamma=0.6}$ &0.008822& -0.002234& -0.012007 &-0.001263& 0.021606&  -0.008041 & -0.009173  & 0.005880\\
$\mathbf{\gamma=0.8}$& 0.009737& -0.001554& -0.010687&-0.001596&0.019549& -0.005746& -0.008246& 0.003201\\
$\mathbf{\gamma=1.0}$ & 0.010302& -0.001089& -0.010195 &-0.001804&0.018889 &-0.004337& -0.007766&  0.002108 \\
\midrule
\multicolumn{9}{l}{$\bm{\Lambda}=\bm{\Lambda}_2$}\\
\midrule
{MLE}&-0.000693&0.010395 &-0.000313  &0.006264 & -0.009748 & 0.031526 & 0.019998 & 0.021093 \\
\midrule
\multicolumn{9}{l}{{WMDPDE}}  \\
\midrule 
$\mathbf{\gamma=0.2}$ &-0.002410  &0.010906 &-0.001018 & 0.007107  &-0.009234&  0.029215&  0.019993& 0.024129\\
$\mathbf{\gamma=0.4}$&-0.002700&  0.014114& -0.001747 & 0.004122&  -0.009129&  0.027226&  0.019995  &0.024251\\
$\mathbf{\gamma=0.6}$&-0.002593&  0.014026& -0.001281 & 0.004645 &-0.009188 & 0.025652&  0.019994&  0.023626 \\
$\mathbf{\gamma=0.8}$&-0.002606&  0.015897& -0.001327& 0.006511&-0.009301 & 0.024423  &0.019994&  0.022890\\
$\mathbf{\gamma=1.0}$&-0.004000&  0.010148&  0.000993 & 0.010087 & -0.009423 & 0.023463&  0.019995 & 0.022233 \\
\bottomrule
\end{tabular}}}}
\label{tab3}
\end{table}
From Table \ref{tab3}, it is evident that the bias of MLE is less than that of WMDPDE in pure data.  When data is contaminated, WMDPDE performs better than MLE as the bias of WMDPDE is less than that of MLE.  Also, the bias change for WMDPDE is lesser than that of MLE, from pure to contamination scheme.  Table \ref{tab4} shows that the bias of BE is less than that of WRBE in the pure data scheme.  However, after contamination, WRBE shows less bias.  Thus, from Tables \ref{tab3} and \ref{tab4}, it can be concluded that WMDPDE and WRBE are robust estimators.  Further, observing the figures in Tables \ref{tab3} and \ref{tab4}, it can be concluded that under contamination, WRBE has the smallest bias compared to other estimates. 
Hence, if prior information is available, WRBE is the best choice, but if it is not possible to get prior knowledge, one can rely on WMDPDE for robust estimation purposes.  Also, absolute bias and RMSE of reliability estimates of parameters for contamination under Sim. 1 setting are represented graphically in Figures \ref{fig1}-\ref{fig3}.  The superiority of robust estimates under contamination is also clearly visible from these figures.
\begin{table}[htb!]
\tbl{Bias of BE and WRBE under simulation.}
{\scalebox{1}{
{\begin{tabular}{lcccccccc} \toprule
&\multicolumn{4}{c}{{Pure Data}} & \multicolumn{4}{c}{{Contamination}} \\\cmidrule{2-9}
& $\hat{a}_1$ & $\hat{b}_1$ & $\hat{a}_2$ & $\hat{b}_2$&  $\tilde{a}_1$ & $\tilde{b}_1$ & $\tilde{a}_2$ & $\tilde{b}_2$\\\midrule
 \multicolumn{9}{c}{{Normal Prior}}\\
 \midrule
 \multicolumn{9}{l}{$\bm{\Lambda}=\bm{\Lambda}_1$}\\
 \midrule
 {BE} & 0.008511 &-0.000794& -0.009001 &-0.001089 &0.019942& -0.001403& -0.011804& -0.003973  \\
 \midrule
 \multicolumn{9}{l}{{WRBE}}  \\
 \midrule
$\mathbf{\gamma=0.2}$ &  0.009226& -0.000977& -0.010004& -0.002014 & 0.009990& -0.000989 &-0.009976&-0.002014 \\
$\mathbf{\gamma=0.4}$&  0.009764& -0.001028& -0.009714& -0.001612 & 0.009985& -0.000989& -0.009996&-0.002013 \\
$\mathbf{\gamma=0.6}$ & 0.009241& -0.001001& -0.009283& -0.001567 & 0.010009& -0.000968& -0.009986&-0.001996\\
$\mathbf{\gamma=0.8}$&  0.010006& -0.001013 &-0.009824 &-0.002014 & 0.009998& -0.000994&-0.009998& -0.001989  \\
$\mathbf{\gamma=1.0}$ & 0.009792& -0.000997& -0.009604& -0.001807 &  0.010025& -0.000981& -0.010016&-0.001973\\
\midrule
\multicolumn{9}{l}{$\bm{\Lambda}=\bm{\Lambda}_2$}\\
\midrule
{BE} &0.000006 &  0.009929& -0.000001  &0.010009 & -0.000073  &0.010103 &0.000020  &0.010031 \\
\midrule
\multicolumn{9}{l}{{WRBE}}  \\
\midrule
$\mathbf{\gamma=0.2}$ & 0.000024 & 0.009955& -0.000013 & 0.009974 & 0.000030 & 0.009968 &0.000010 &0.009999 \\
$\mathbf{\gamma=0.4}$&-0.000008 &0.009980& 0.000002 &  0.009982&  0.000014&  0.010028  &-0.000004 & 0.010001 \\
$\mathbf{\gamma=0.6}$ & 0.000009 &0.009979& 0.000004&0.009969 & 0.000060&0.009988  &0.000006  &0.009983 \\
$\mathbf{\gamma=0.8}$& 0.000021 &  0.009938& -0.000013&0.009974   & -0.000034&  0.009945&-0.000018& 0.010008  \\
$\mathbf{\gamma=1.0}$ & 0.000028 &0.010017 &    -0.000006&0.009972  &-0.000047 &0.010039  &0.000026  &0.010022\\
\midrule
 \multicolumn{9}{c}{{Dirichlet Prior}}\\
 \midrule
 \multicolumn{9}{l}{$\bm{\Lambda}=\bm{\Lambda}_1$}\\
 \midrule
 {BE} &  0.009434& -0.001010& -0.010029& -0.001008& 0.012976& -0.001770& -0.010445& -0.002651\\
 \midrule
 \multicolumn{9}{l}{{WRBE}}  \\
 \midrule
 $\mathbf{\gamma=0.2}$ & 0.009817 &-0.000938& -0.011580& -0.001155 & 0.010042 &-0.001057 &-0.009628& -0.001896 \\
$\mathbf{\gamma=0.4}$& 0.009707 &-0.000934& -0.010160 &-0.001812 &0.010488& -0.001133 &-0.010255& -0.001835 \\
$\mathbf{\gamma=0.6}$ & 0.009784& -0.001042& -0.009966 &-0.002031 &0.009584& -0.001216 &-0.010399& -0.002060  \\
$\mathbf{\gamma=0.8}$&  0.009814& -0.001001& -0.009879 &-0.002003& 0.009545& -0.000989 &-0.009802& -0.002114 \\
$\mathbf{\gamma=1.0}$ & 0.009836& -0.001000 &-0.010015 &-0.002000&0.010028& -0.001007 &-0.010238 &-0.002139\\
\midrule
\multicolumn{9}{l}{$\bm{\Lambda}=\bm{\Lambda}_2$}\\
\midrule
{BE} & 0.000004  &0.009158 &-0.000019 &  0.009561& -0.000077&0.012739 &-0.000055 & 0.010048 \\
\midrule
\multicolumn{9}{l}{{WRBE}}  \\
\midrule
$\mathbf{\gamma=0.2}$ &-0.000029  & 0.009476 & -0.000019&0.010011 &-0.000056 &  0.010023&0.000026  &  0.009876 \\
$\mathbf{\gamma=0.4}$&-0.000047 &  0.009306 &-0.000025 & 0.010022 &-0.000053& 0.010013  &  0.000026 & 0.010028 \\
$\mathbf{\gamma=0.6}$ &  -0.000020 & 0.009807 &0.000001&  0.010011& -0.000046 & 0.009661&-0.000009  & 0.009891 \\
$\mathbf{\gamma=0.8}$& -0.000033 & 0.009550  & 0.000004 &0.010007  & -0.000060 & 0.009721 &0.000019 & 0.010031\\
$\mathbf{\gamma=1.0}$ & -0.000044  & 0.010016  &0.000008&  0.009896 &-0.000041 &  0.009893 &-0.000008&  0.010004\\
\bottomrule
\end{tabular}}}}
\label{tab4}
\end{table}

\subsection{Influence function analysis}
\begin{figure}[htb!]
\centering
\subfloat[\textrm{\scriptsize{$a_1$ with $\gamma=0.2$}}]{\includegraphics[height=3.4cm,width =0.28\textwidth]{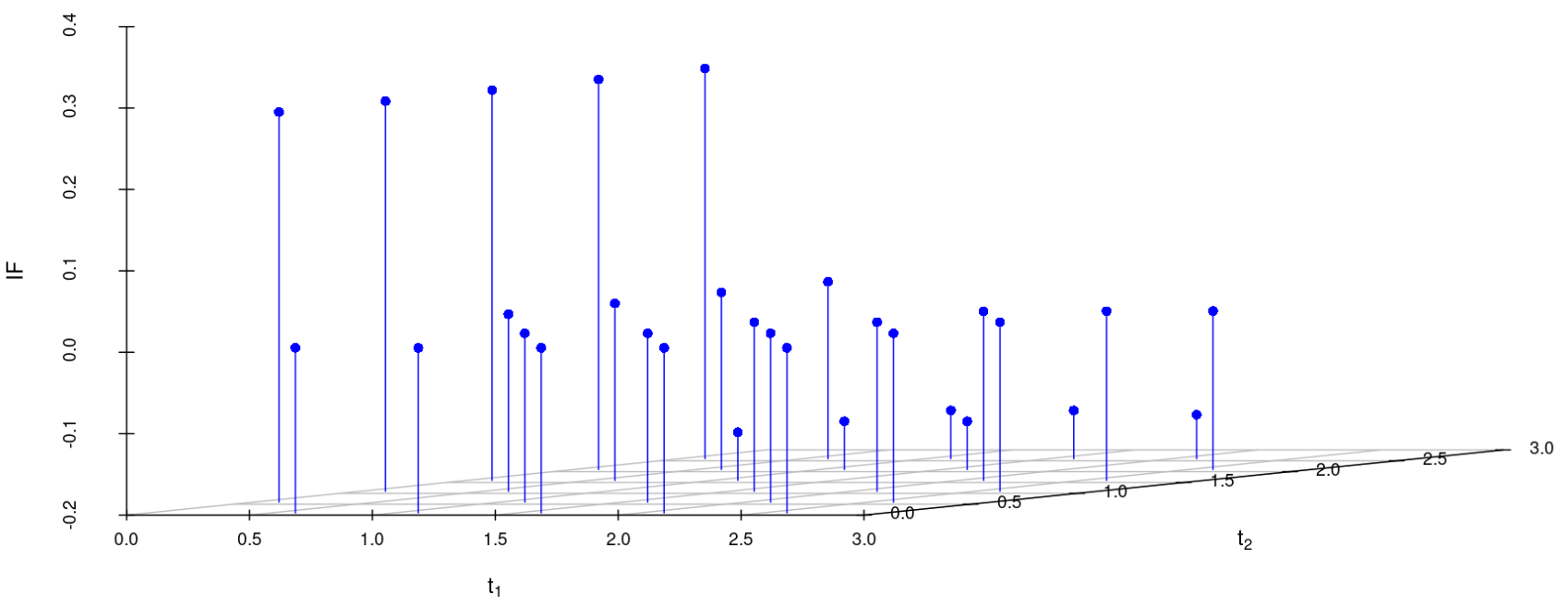}} 
\subfloat[\textrm{\scriptsize{$b_1$ with $\gamma=0.2$}}]{\includegraphics[height=3.4cm,width =0.28\textwidth]{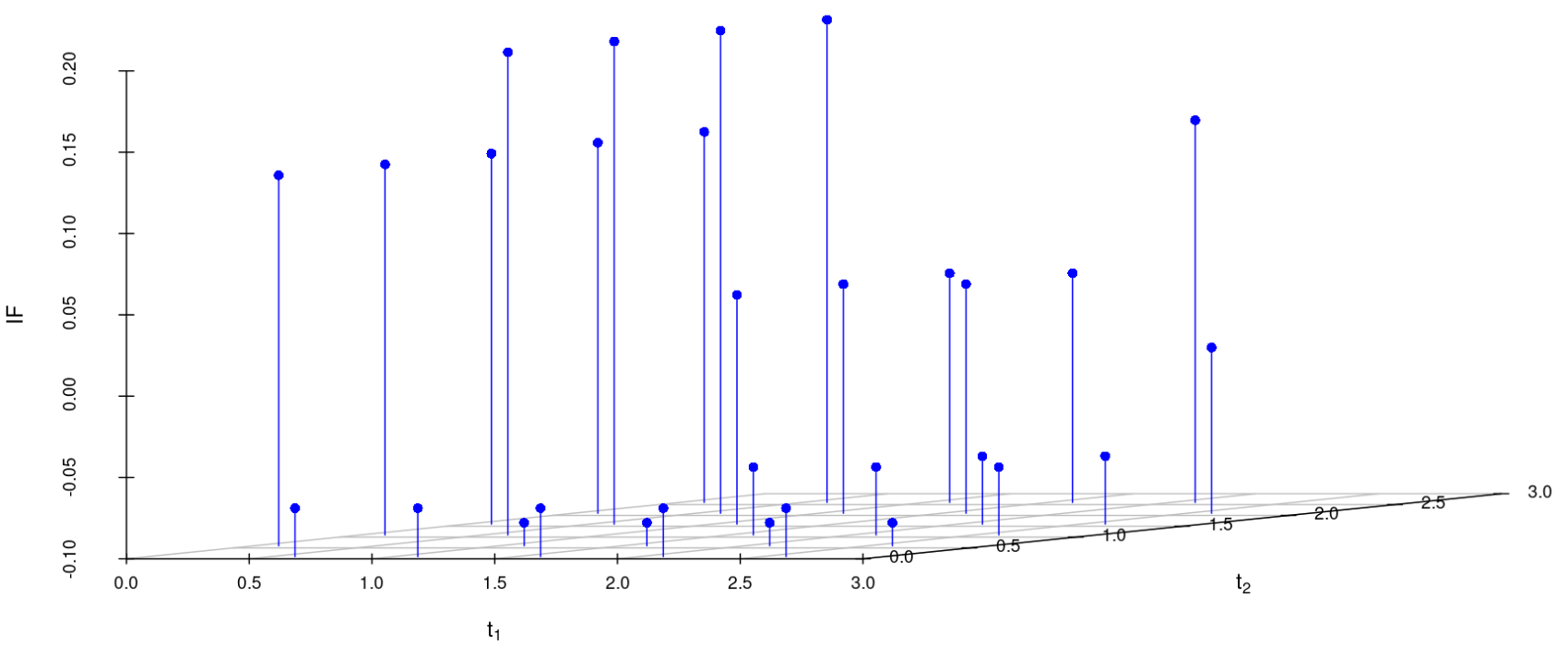}}
\subfloat[\textrm{\scriptsize{$a_2$ with $\gamma=0.2$}}]{\includegraphics[height=3.4cm,width =0.28\textwidth]{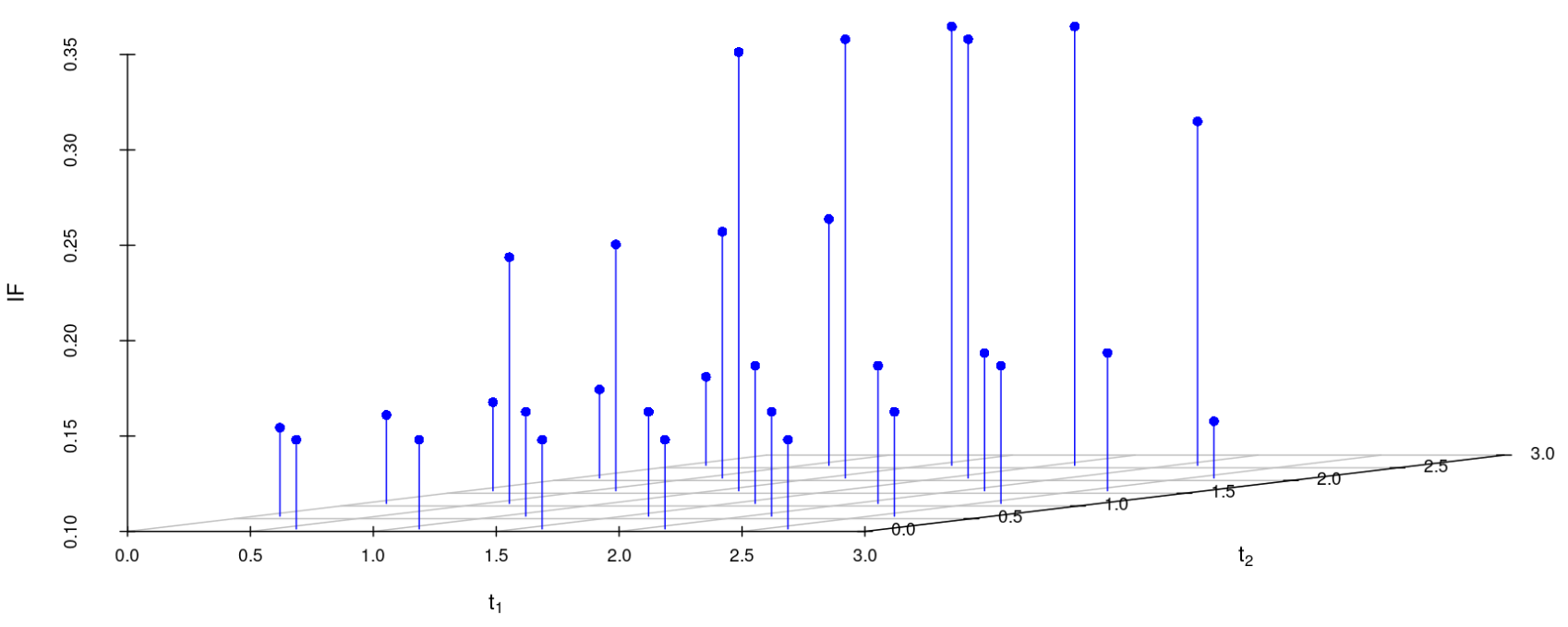}}
\subfloat[\textrm{\scriptsize{$b_2$ with $\gamma=0.2$}}]{\includegraphics[height=3.4cm,width =0.28\textwidth]{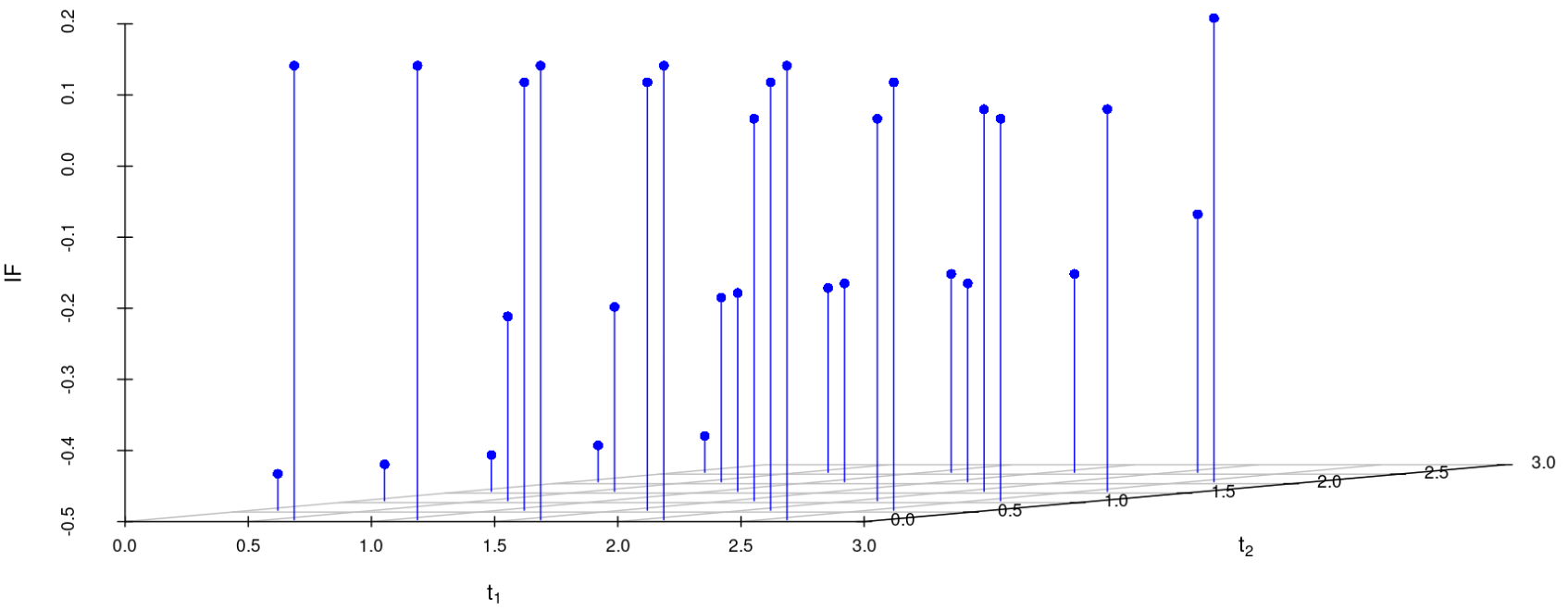}}\\
\subfloat[\textrm{\scriptsize{$a_1$ with $\gamma=0.8$}}]{\includegraphics[height=3.4cm,width =0.28\textwidth]{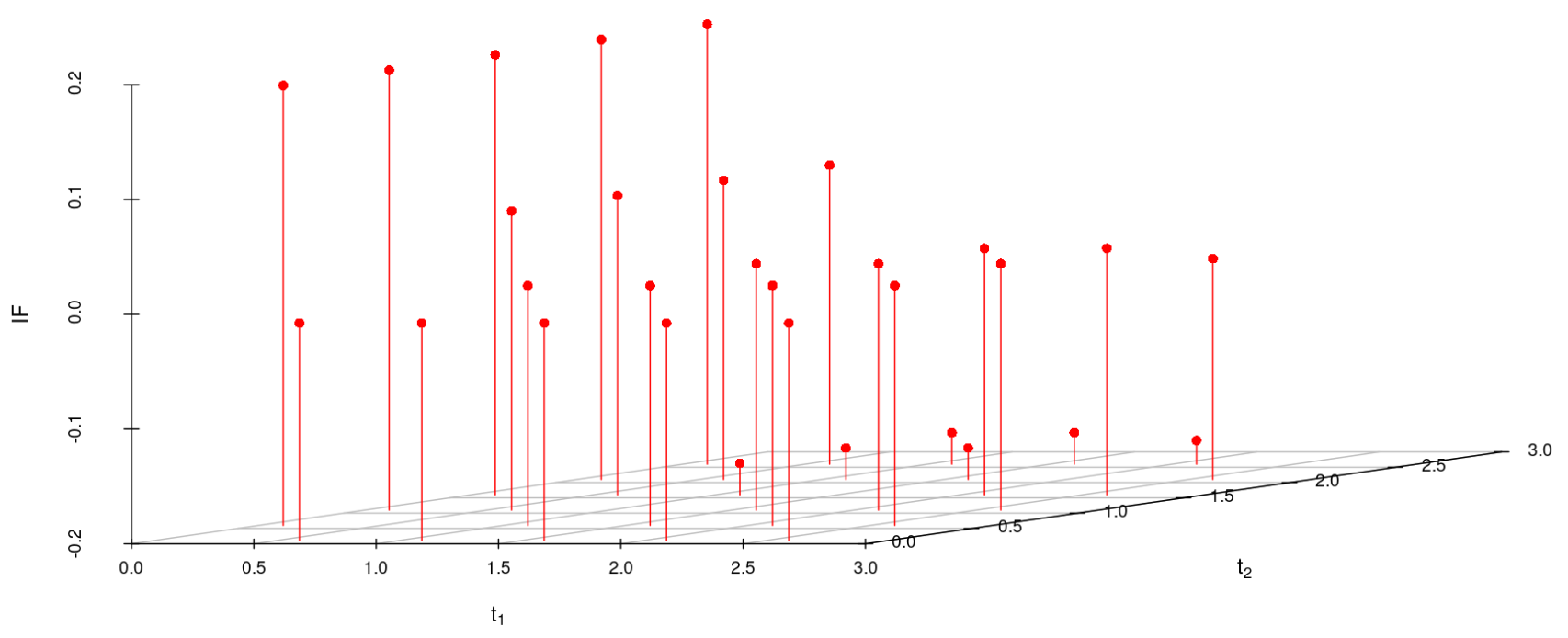}}
\subfloat[\textrm{\scriptsize{$b_1$ with $\gamma=0.8$}}]{\includegraphics[height=3.4cm,width =0.28\textwidth]{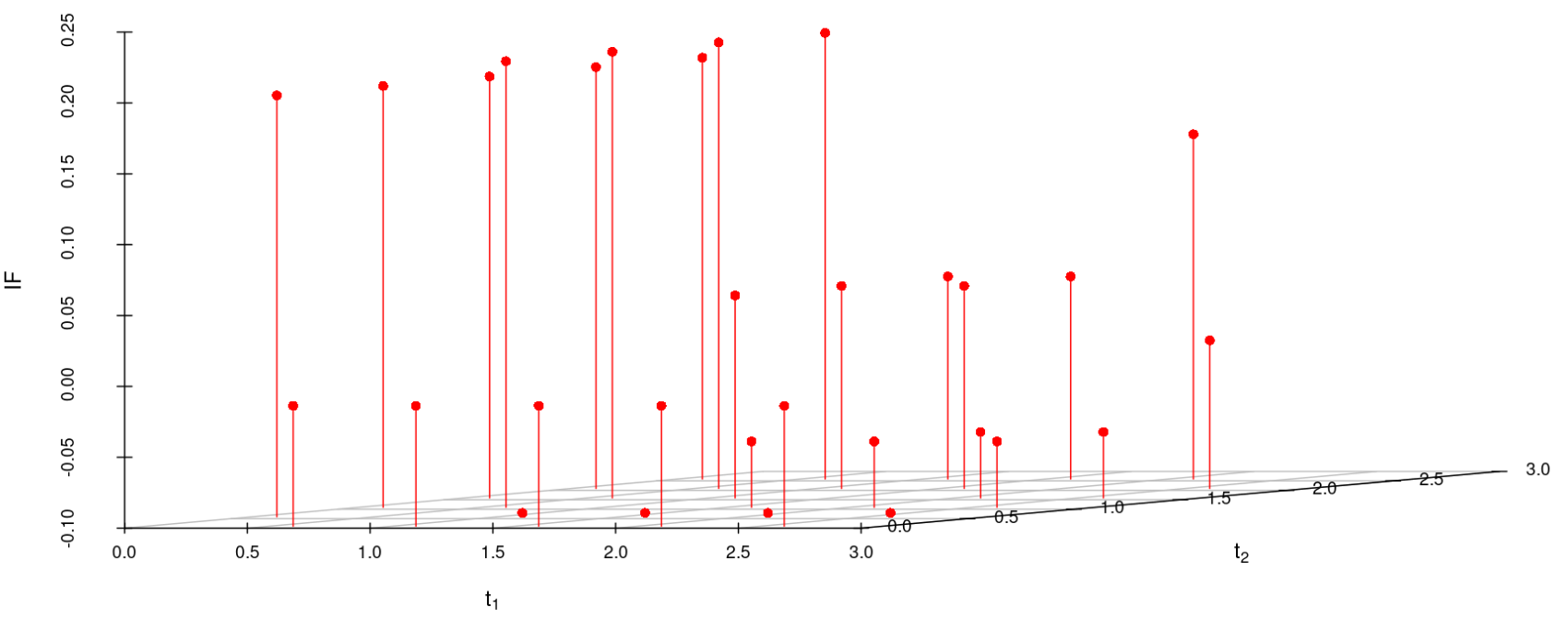}}
\subfloat[\textrm{\scriptsize{$a_2$ with $\gamma=0.8$}}]{\includegraphics[height=3.4cm,width =0.28\textwidth]{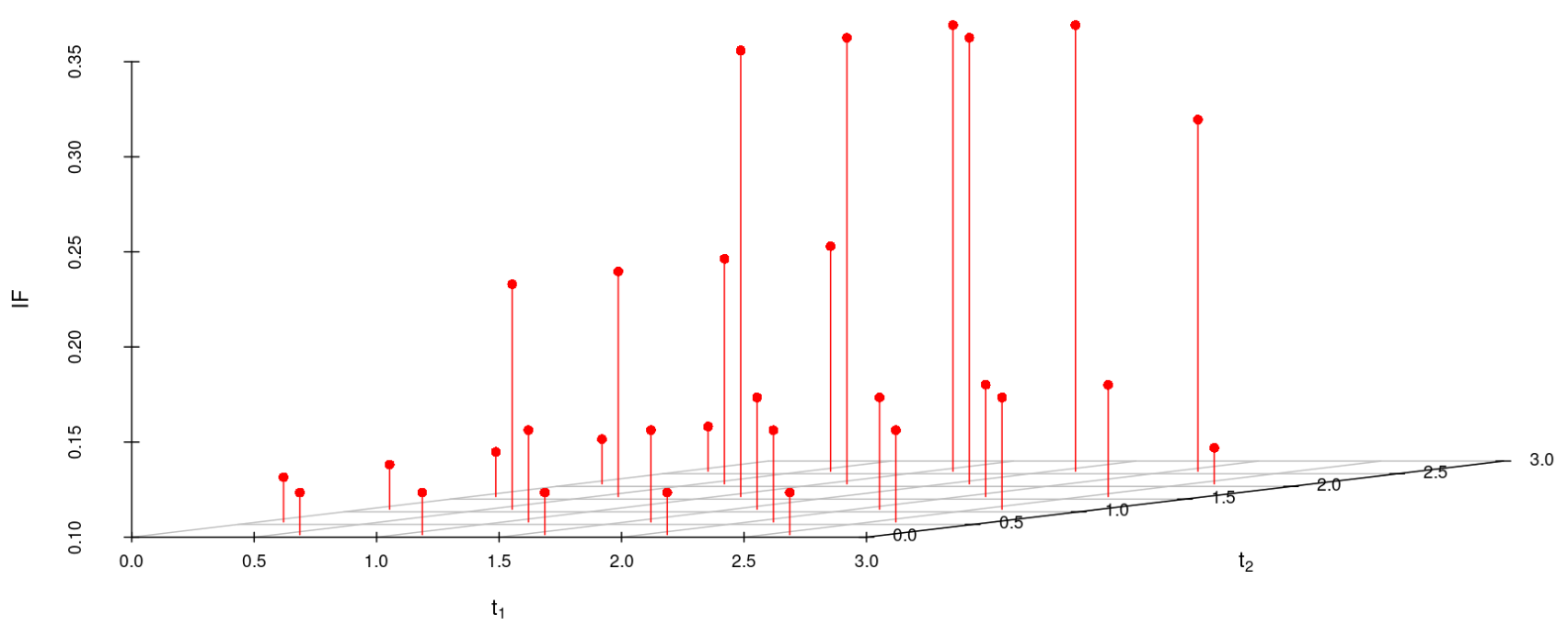}}
\subfloat[\textrm{\scriptsize{$b_2$ with $\gamma=0.8$}}]{\includegraphics[height=3.4cm,width =0.28\textwidth]{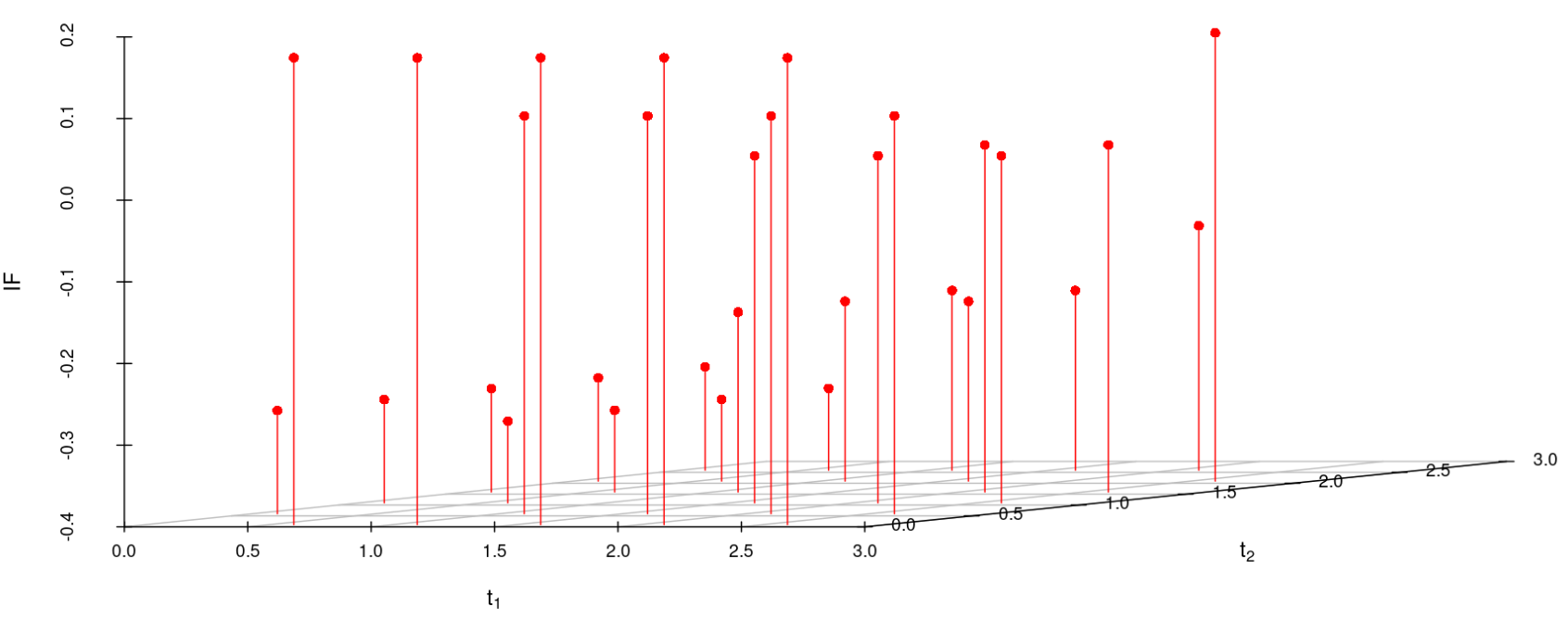}}
\caption{Influence function of WMDPDE.}
\label{fig4}
\end{figure}

\begin{figure}[htb!]
\centering
\subfloat[\textrm{\scriptsize{$a_1$ with $\gamma=0.2$}}]{\includegraphics[height=3.4cm,width =0.28\textwidth]{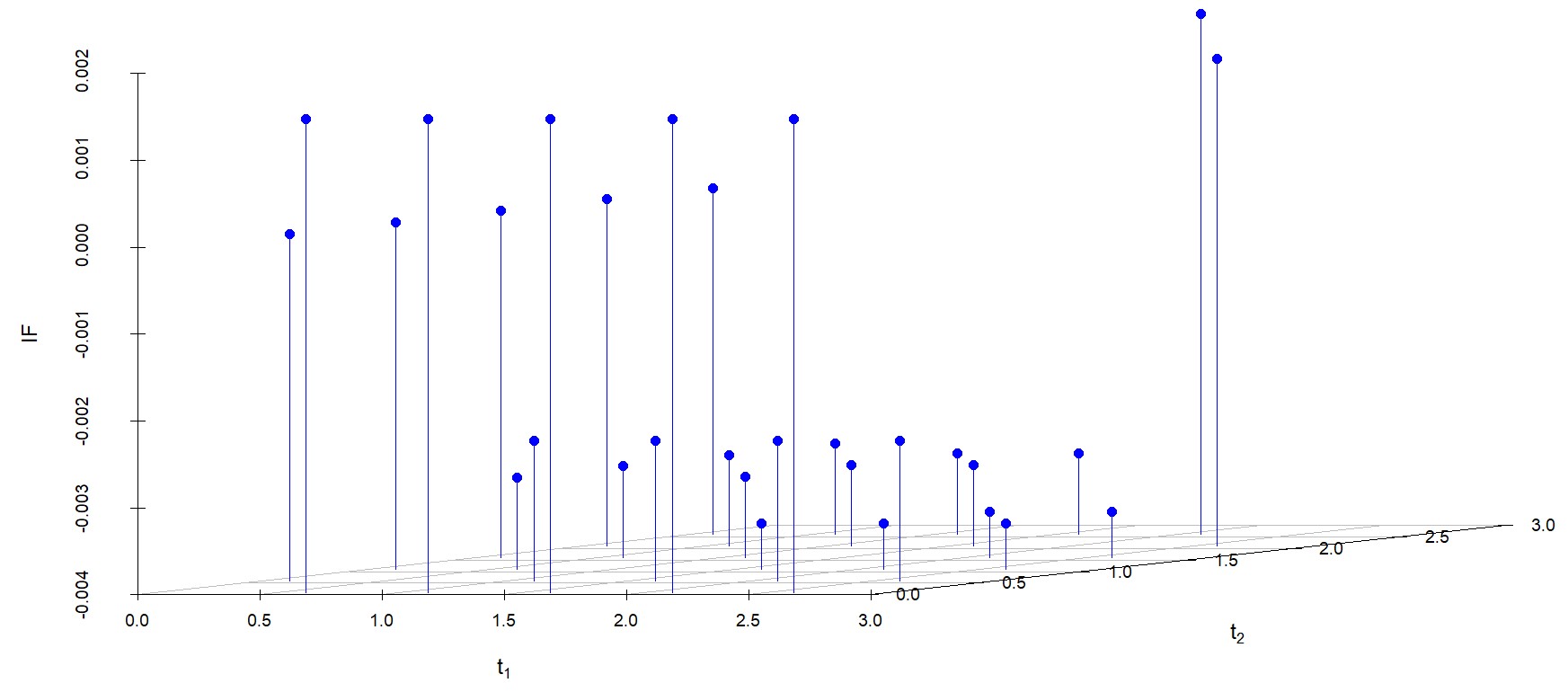}} 
 \subfloat[\textrm{\scriptsize{$b_1$ with $\gamma=0.2$}}]{\includegraphics[height=3.4cm,width =0.28\textwidth]{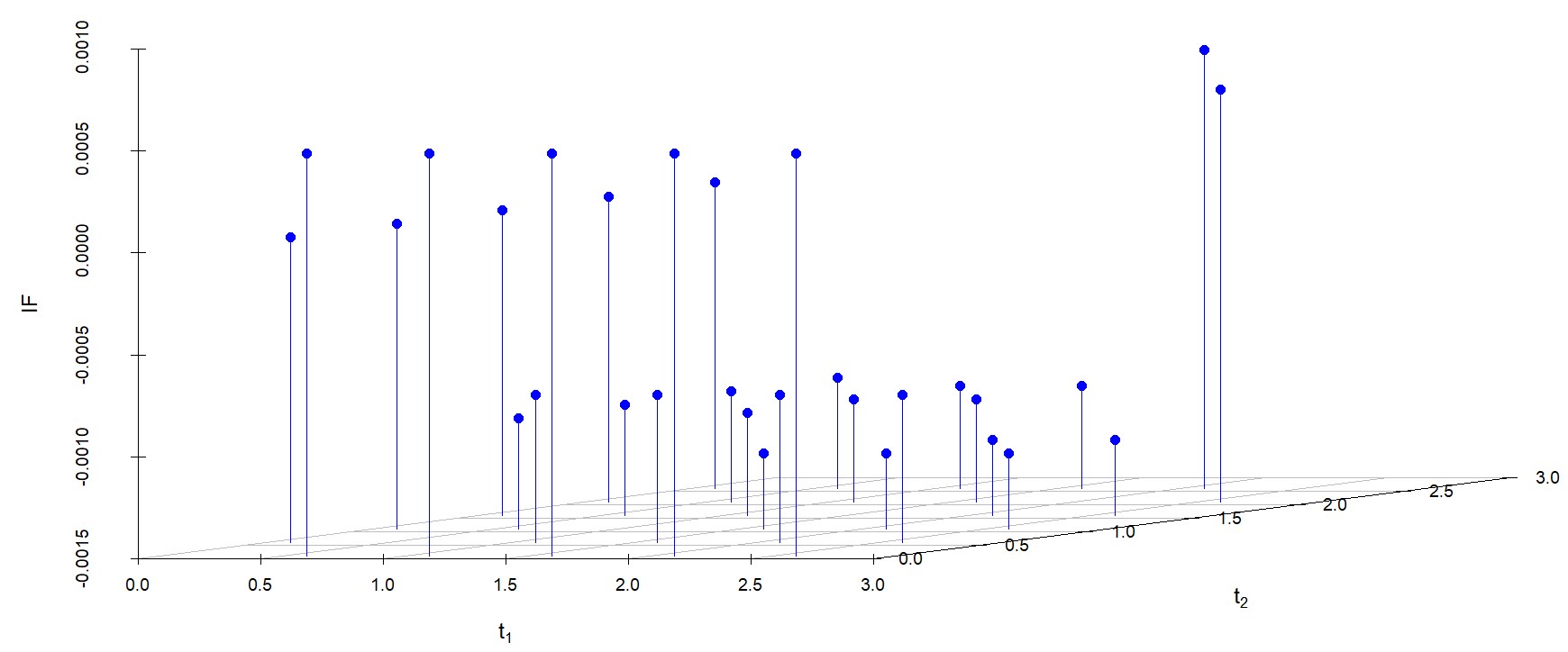}}
 \subfloat[\textrm{\scriptsize{$a_2$ with $\gamma=0.2$}}]{\includegraphics[height=3.4cm,width =0.28\textwidth]{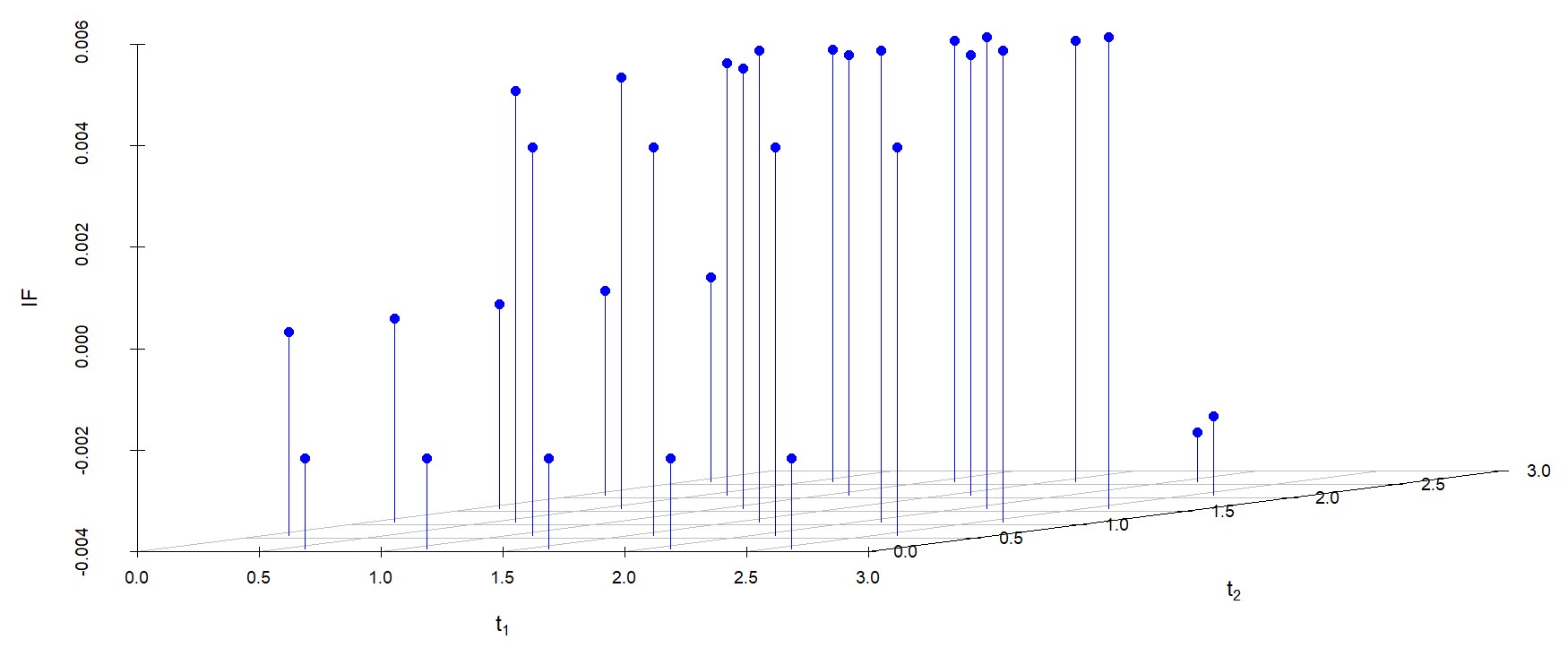}}
\subfloat[\textrm{\scriptsize{$b_2$ with $\gamma=0.2$}}]{\includegraphics[height=3.4cm,width =0.28\textwidth]{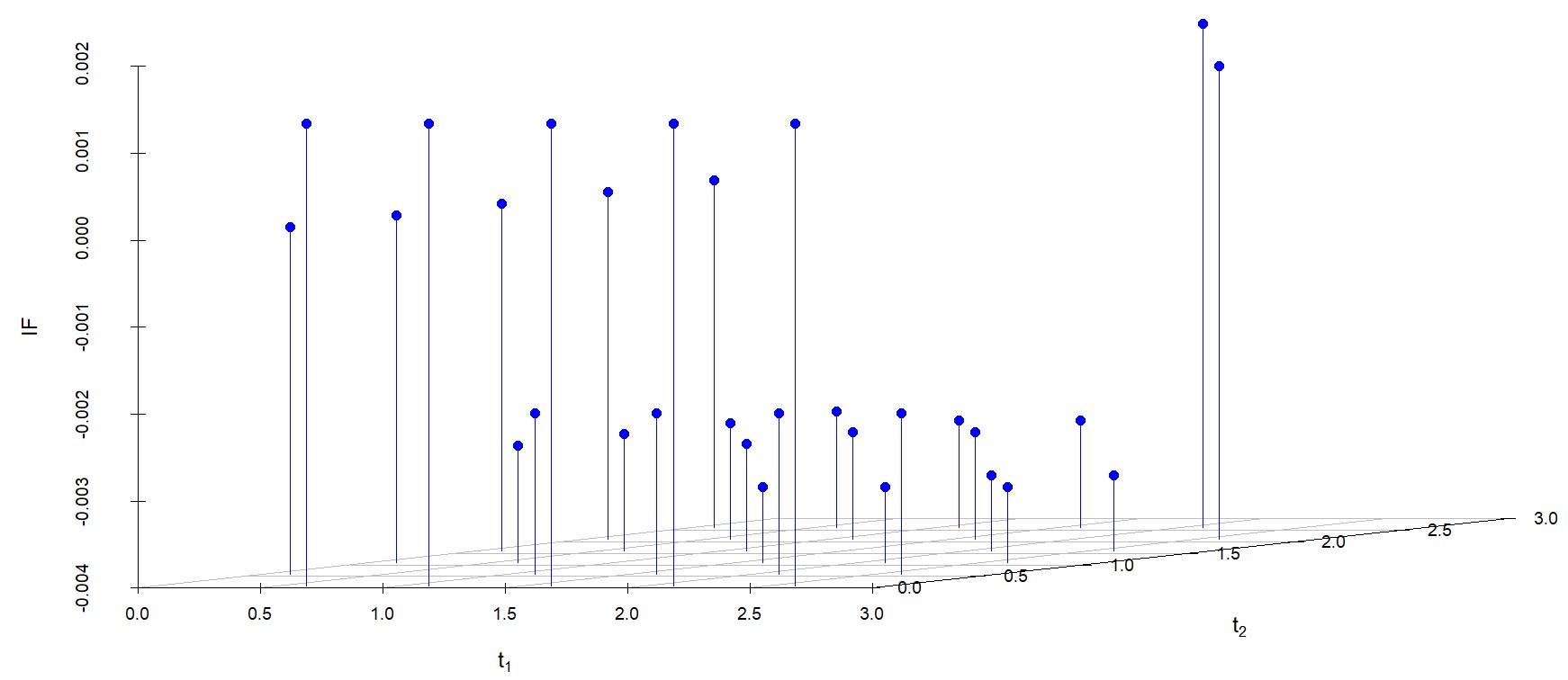}}\\
\subfloat[\textrm{\scriptsize{$a_1$ with $\gamma=0.8$}}]{\includegraphics[height=3.4cm,width =0.28\textwidth]{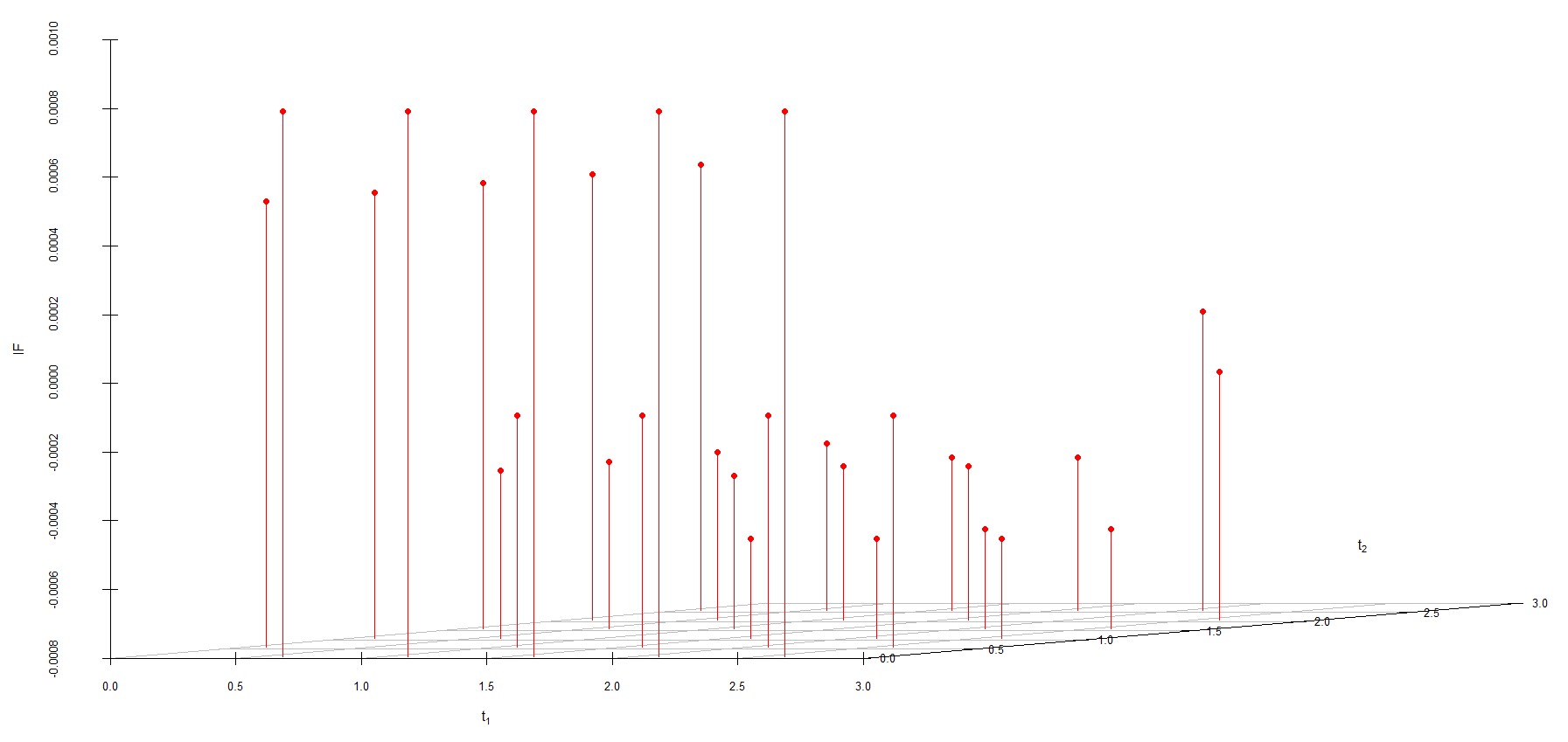}}
\subfloat[\textrm{\scriptsize{$b_1$ with $\gamma=0.8$}}]{\includegraphics[height=3.4cm,width =0.28\textwidth]{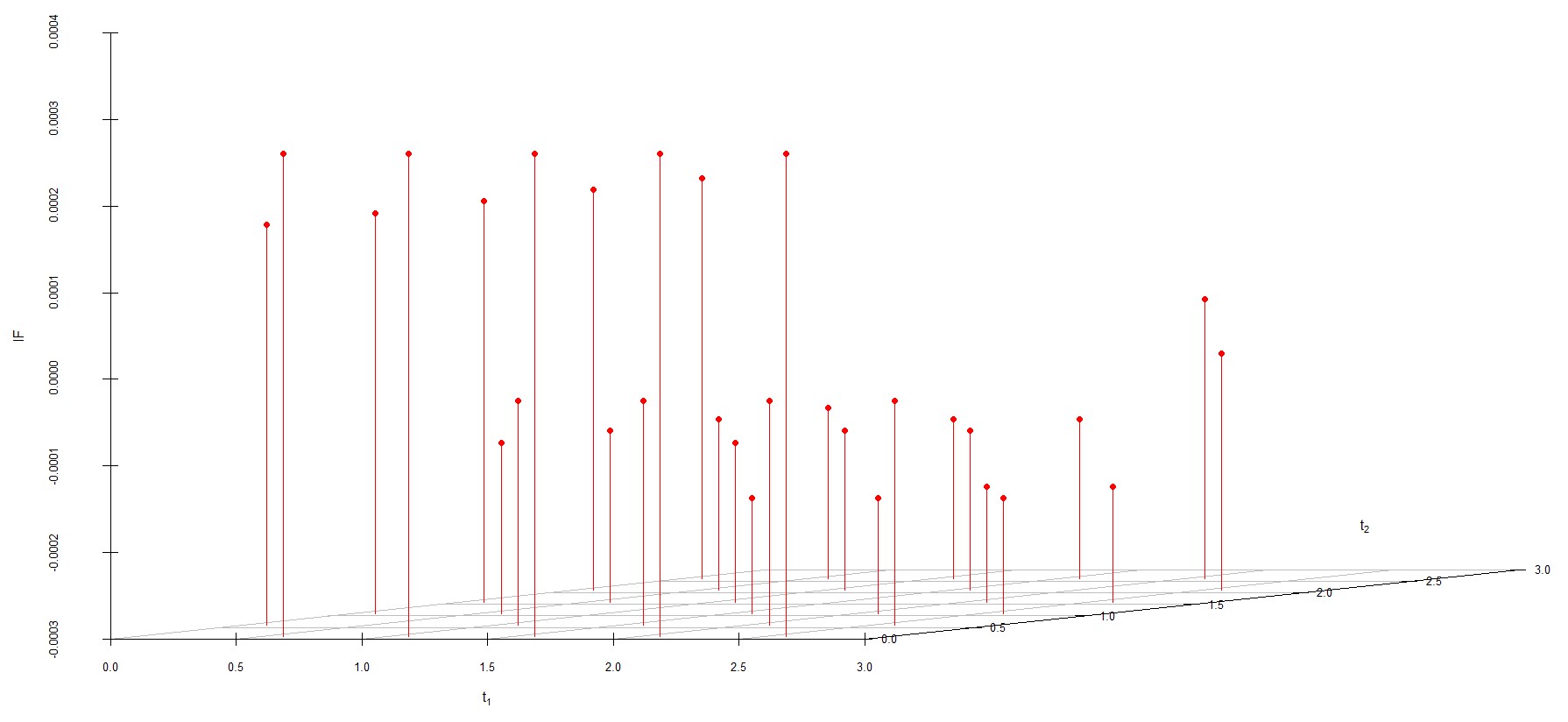}} 
\subfloat[\textrm{\scriptsize{$a_2$ with $\gamma=0.8$}}]{\includegraphics[height=3.4cm,width =0.28\textwidth]{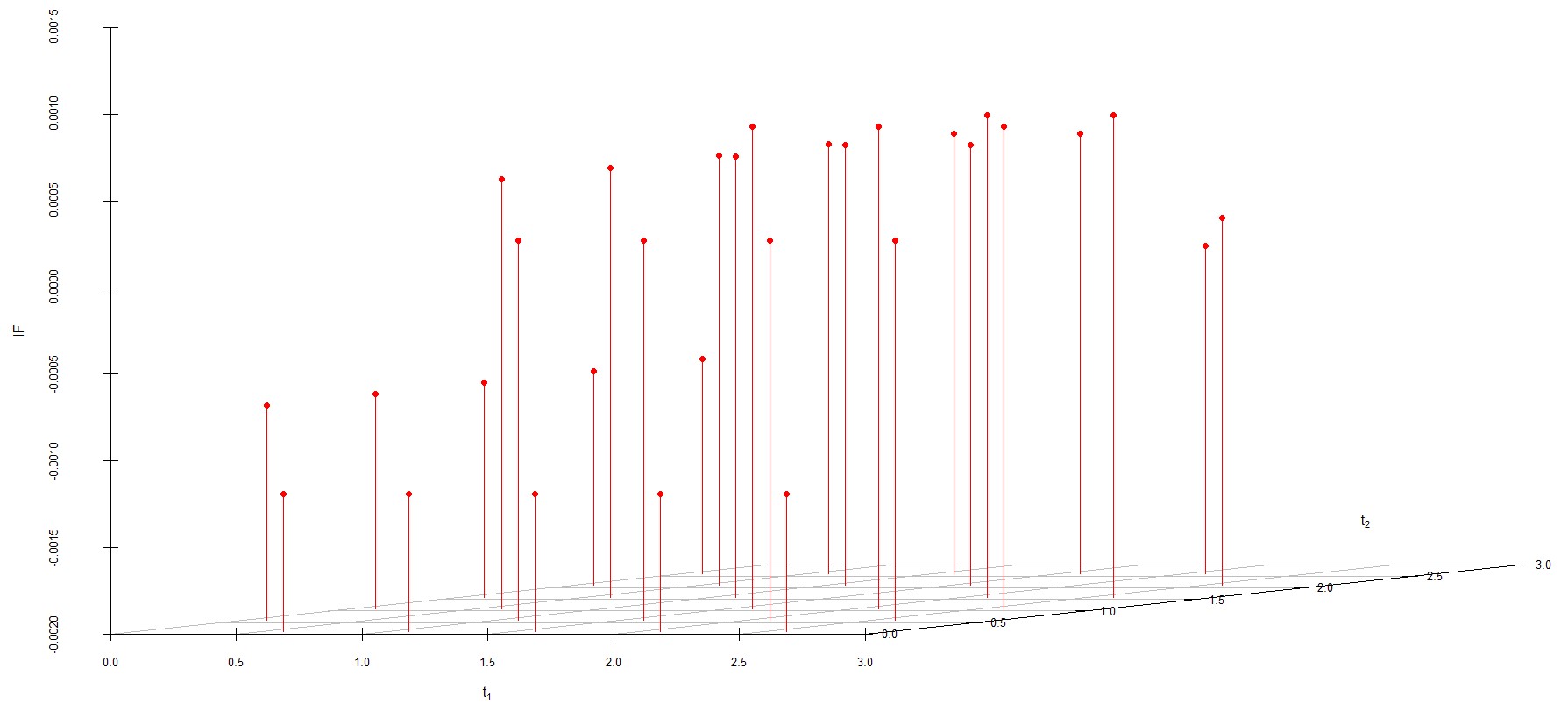}}
\subfloat[\textrm{\scriptsize{$b_2$ with $\gamma=0.8$}}]{\includegraphics[height=3.4cm,width =0.28\textwidth]{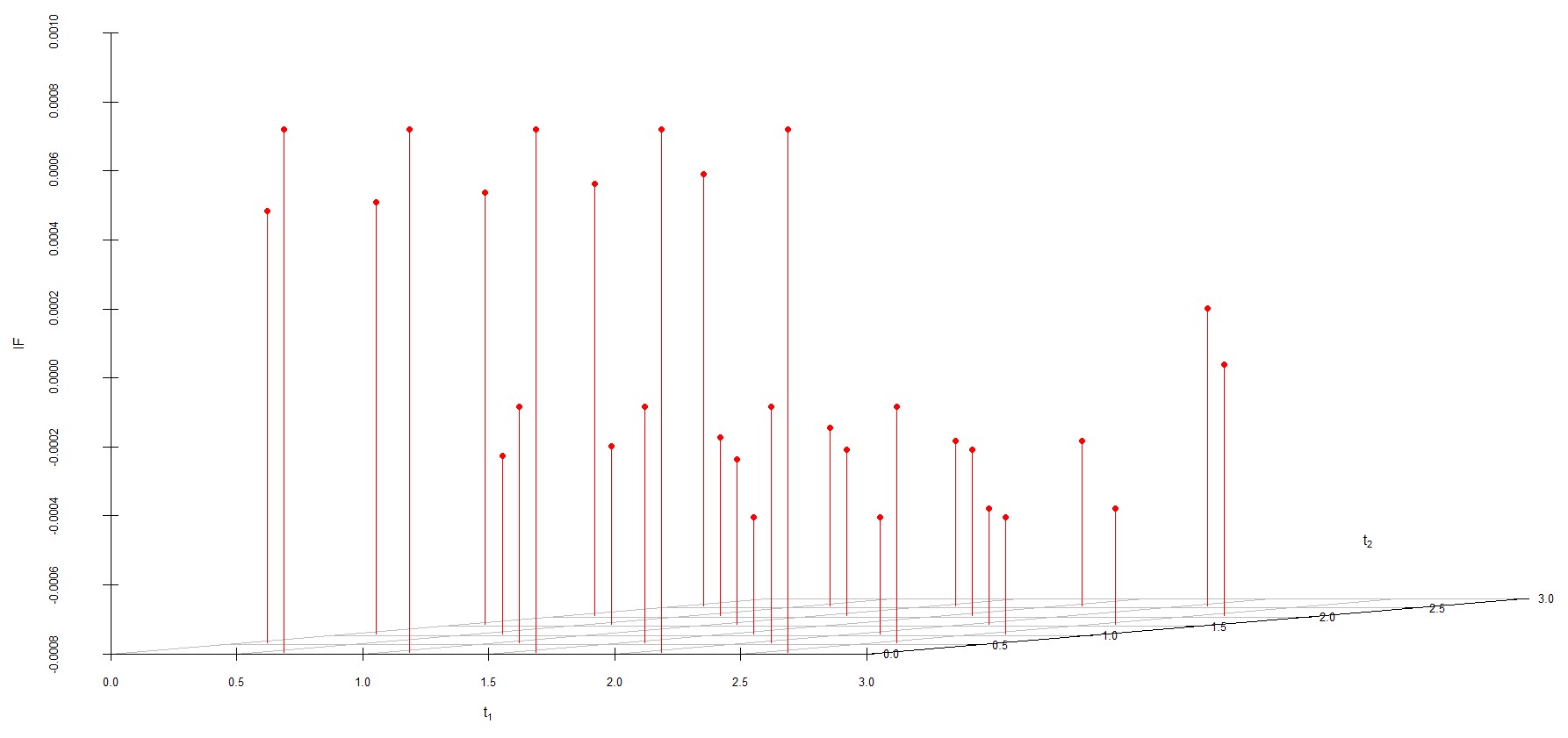}}
\caption{Influence function of WRBE (Normal prior).}
\label{fig5}
\end{figure}

We study the influence of WMDPDE and WRBE as described in section \eqref{sec5}.  The influence functions are empirically computed for Sim. 1 with tuning parameters $\gamma=(0.2, 0.8)$ and are represented through the Figures \ref{fig4}-\ref{fig5}.  From these figures, it has been observed that the influence function is bounded, which verifies the robustness of the estimates.  Since the bounds for WRBE under normal and Dirichlet prior came out to be similar, the IF figure for 
\noindent Dirichlet prior is omitted here.  Further, the influence function values for WRBE are less than those of WMDPDE for all the parameters.  When the tuning parameter value is increased from $\gamma=0.2$ to $\gamma=0.8$, the influence function values for WRBE decrease.  Thus, the robustness of WRBE is increased with an increased value of $\gamma$.

\subsection{Data Analysis}
The application of NOSD test data can be extended to the survival analysis of biomedical data.  Therefore, data for the analysis is extracted from the SEER research database (http://www.seer.cancer.gov). 
 The information in this extracted dataset is focused only on patients diagnosed with pancreatic cancer during the year 2016 in the age group 50-55.  A total of 235 patients are involved in the study where patient death due to pancreatic cancer is taken as competing cause 1, and death due to other causes is taken as competing cause 2.  The size of the tumour is taken as stress level, which is indicated as,
$$
x_i=\begin{cases}
1, & \text{if tumor size}\leq 30\,mm\\
2, & \text{if}\quad 30\,mm< \text{tumor size}\leq 45\,mm\\
3, & \text{if tumor size}\geq 45\,mm
		 \end{cases}.
$$ 
Patients are observed over the months, and deaths are recorded within the follow-up time.  The data layout is described in Table \ref{tab5}.
\begin{table}[htb!]
\tbl{Layout of the SEER pancreatic cancer data (www.seer.cancer.gov).}
{\scalebox{1}{
{\begin{tabular}{lccccccccc} \toprule
{Groups}&{Diagnosed} & {Tumor}& \multicolumn{3}{c}{{Inspection Times}}&\multicolumn{3}{c}{{Deaths}}&Survived\\
&{patients} &{size} &\multicolumn{3}{c}{{(Months)}}&\multicolumn{3}{c}{{(Cancer, Other)}}&Patients \\\midrule
1 & 69 & 1 & 02 & 10 & 30&(07,1) & (26,0) & (28,2) &5 \\
2 & 90 & 2 & 01 & 10 & 34 & {(14,1)} & (33,1)   & (31,3)&7 \\
3 & 76 & 3 & 01 & 08 & 20 & {(21,1)} & {(23,1)}   & (22,1)&7\\
\bottomrule
\end{tabular}}}}
\label{tab5}
\end{table}
A bootstrap goodness of fit test is performed to ensure that the two-parameter Lindley distribution is fitted to the data.  The distance-based test statistic is given as, 
$$T=max\vert \tilde{q}_{ilr}-\hat{p}_{ilr}\vert\;,l=1,2,3\,,r=0,1,2,$$ where, $(.)_{il0}=(.)_{i0}$, $\tilde{q}_{(.)}$, $\hat{p}_{(.)}$ are the empirical and estimated failure or survival probability respectively.  For estimation purposes, the initial values found through the grid-search procedure are $(a1{=}{-}0.6, b1{=}0.34, a2{=}0.5, b2{=}0.1)$.  MLE given in Table \ref{tab6} is used to obtain the expected number of deaths and survivals.  The value of the test statistic came out to be $0.388285$, and the corresponding approximate p-value is $0.528$, which strongly satisfies the assumption of Lindley distribution as the lifetime distribution.  The estimates derived from MLE and WMDPDE with $95\%$ asymptotic confidence interval (CI) are given in Tabel \ref{tab6}.  Similarly, Bayes estimates (BE) and weighted robust Bayes estimates (WRBE) with $95\%$ Highest posterior density credible interval (HPD CRI) are reported in Table \ref{tab7}.  The bootstrap bias (BT Bias) and root mean square of error (RMSE) of the estimates are shown in Table \ref{tab8}.  It can be seen from this table that the BT bias and RMSE of robust estimates are less than those of classical estimates.  It is also observed that WRBE has the least bias and RMSE among the four estimation methods, which conforms to the findings of the simulation experiment and aligns with the objective of this study. 
\begin{table}[htb!]
\tbl{Classical estimates (95\% Asymp. CI) for real data.}
{\scalebox{1}{
{\begin{tabular}{lcccc} \toprule
 &$\hat{\bm{a}}_1$&$\hat{\bm{b}}_1$&$\hat{\bm{a}}_2$&$\hat{\bm{b}}_2$\\ \cmidrule{2-5}
&{Est. (CI)}&{Est. (CI)}&{Est. (CI)}&{Est. (CI)}\\ \midrule
{MLE} & -0.611204&  0.198243& 0.399344&  0.162147 \\
&(-0.6249,-0.5974) & (0.1951,0.2013) & (0.3801,0.4185) & (0.1547,0.1695)\\\midrule
\multicolumn{5}{l}{{WMDPDE}}  \\\midrule
$\mathbf{\gamma=0.2}$ & -0.611665 & 0.087949&  0.440619  &0.209047  \\
&(-0.6225,-0.6007)&(0.0852,0.0906)&  (0.4184,0.4627) & (0.2016,0.2164)\\
$\mathbf{\gamma=0.4}$& -0.600050 & 0.339492 & 0.499800 & 0.100896 \\
&(-0.6142,-0.5858)&  (0.3363,0.3426) & (0.4854,0.5141) & (0.0939,0.1078)\\
$\mathbf{\gamma=0.6}$ & -0.600021 & 0.339811 & 0.499956& 0.100701\\
&(-0.6133,-0.5866) & (0.3366,0.3429)&  (0.4812,0.5186) & (0.0934,0.1079)\\
$\mathbf{\gamma=0.8}$& -0.600009 & 0.339983&  0.499996 & 0.100549 \\
&(-0.6138,-0.5861) & (0.3367,0.3431)&  (0.4674,0.5325) & (0.0928,0.1082)\\
$\mathbf{\gamma=1.0}$ &-0.600002 & 0.340070 & 0.500005 & 0.100431 \\
& (-0.6180,-0.5819) & (0.3367,0.3433) & (0.4311,0.5689) & (0.0917,0.1090)\\  \bottomrule
\end{tabular}}}}
\label{tab6}
\end{table}

\begin{table}[htb!]
\tbl{Bayes estimates  (95\% HPD CRI) for real data.}
{\scalebox{1}{
{\begin{tabular}{lcccc} \toprule
&$\hat{\bm{a}}_1$&$\hat{\bm{b}}_1$&$\hat{\bm{a}}_2$&$\hat{\bm{b}}_2$\\ \cmidrule{2-5}
&{Est. (CI)}&{Est. (CI)}&{Est. (CI)}&{Est. (CI)}\\ \midrule
\multicolumn{5}{c}{{Normal prior}}\\ \midrule
{BE} & -0.600302 & 0.339304&  0.499718&  0.099922 \\
&(-0.6049,-0.5958) & (0.3341,0.3433) & (0.4945,0.5045)& (0.0952,0.1041)\\
\midrule
\multicolumn{5}{l}{{WRBE}}  \\
\midrule
$\mathbf{\gamma=0.2}$ &  -0.599938&  0.340171 & 0.500442  &0.099048  \\
&(-0.6015,-0.5984) & (0.3385,0.3416) & (0.4988,0.5020) & (0.0972,0.1006)\\
$\mathbf{\gamma=0.4}$& -0.599988 & 0.339434 & 0.500200 & 0.100394 \\
 &(-0.6014,-0.5984) & (0.3378,0.3409) & (0.4986,0.5016) & (0.0989,0.1019)\\
$\mathbf{\gamma=0.6}$ &-0.599946 & 0.339543&  0.499693&  0.099116 \\
&(-0.6016,-0.5985) & (0.3379,0.3412) & (0.4980,0.5013)  &(0.0975,0.1006)\\
$\mathbf{\gamma=0.8}$&-0.600584&  0.340366 & 0.499820 & 0.100391  \\
&(-0.6019,-0.5986) & (0.3386,0.3419) & (0.4981,0.5012)&  (0.0985,0.1018)\\
$\mathbf{\gamma=1.0}$&-0.599790 & 0.340013 & 0.500456 & 0.100688  \\
&(-0.6012,-0.5981) & (0.3385,0.3416) & (0.4989,0.5021)  &(0.0992,0.1022)\\
\midrule
\multicolumn{5}{c}{{Dirichlet prior}}\\
\midrule
{BE} &  -0.600644 & 0.340258 & 0.499893 & 0.100295\\
&(-0.6052,-0.5958) & (0.3356,0.3446) & (0.4955,0.5052)  &(0.0953,0.1047)\\
\multicolumn{5}{l}{{WRBE}}  \\
\midrule 
$\mathbf{\gamma=0.2}$ & -0.599988 & 0.339434 & 0.500200 & 0.100394  \\
&(-0.6014,-0.5984)&  (0.3378,0.3409)&  (0.4986,0.5016)&(0.0989,0.1019)\\
$\mathbf{\gamma=0.4}$& -0.599946 & 0.339543 & 0.499693 & 0.099116  \\
&(-0.6016,-0.5985)&  (0.3379,0.3412)&  (0.4980,0.5013) & (0.0975,0.1006)\\
$\mathbf{\gamma=0.6}$ &-0.600584 & 0.340366&  0.499820&  0.100391 \\
&(-0.6019,-0.5986) & (0.3386,0.3419)&  (0.4981,0.5012) & (0.0985,0.1018)\\
$\mathbf{\gamma=0.8}$&-0.599790 & 0.340013&  0.500456&  0.100688  \\
&(-0.6012,-0.5981) & (0.3385,0.3416)&  (0.4989,0.5021)&  (0.0992,0.1022)\\
$\mathbf{\gamma=1.0}$&-0.600246&  0.340003&  0.499874 & 0.099856  \\
&(-0.6016,-0.5986)&  (0.3384,0.3417)&  (0.4983,0.5013) & (0.0981,0.1014)\\\bottomrule
\end{tabular}}}}
\label{tab7}
\end{table}

\begin{table}[htb!]
\tbl{BT bias and RMSE of the estimates for real data.}
{\scalebox{1}{
{\begin{tabular}{lcccccccc} \toprule
&\multicolumn{2}{c}{$\hat{\bm{a}}_1$} & \multicolumn{2}{c}{$\hat{\bm{b}}_1$} &\multicolumn{2}{c}{$\hat{\bm{a}}_2$} & \multicolumn{2}{c}{$\hat{\bm{b}}_2$} \\ \cmidrule{2-9}
&{BT Bias}& {RMSE}&{BT Bias}& {RMSE}&{BT Bias}& {RMSE} &{BT Bias}& {RMSE}\\\midrule
 {MLE} &  -0.011560 & 0.012516& -0.063412&  0.075775 &-0.056070&  0.062076 & 0.003658&  0.004223 \\
 \midrule
 \multicolumn{9}{l}{{WMDPDE}}\\
 \midrule
$\mathbf{\gamma=0.2}$&-0.005313&  0.005974& -0.030313&  0.035592& -0.017427&  0.019746&  0.002893&  0.003343\\
$\mathbf{\gamma=0.4}$&-0.002404&  0.002756& -0.013876&  0.016206& -0.004721&  0.005412&  0.002260& 0.002611\\
$\mathbf{\gamma=0.6}$&-0.001099&  0.001270& -0.005113&  0.005961&-0.001065&  0.001227&  0.001763& 0.002036\\
$\mathbf{\gamma=0.8}$&-0.000460&  0.000533& -0.000489&  0.000576& -0.000092  &0.000107&0.001377 & 0.001590 \\
$\mathbf{\gamma=1.0}$&-0.000125 & 0.000144&  0.001814&  0.002101&  0.000139&  0.000161&  0.001078&0.001245\\
\midrule
\multicolumn{9}{c}{{Normal Prior}}\\
\midrule
{BE} & -0.000862&0.000950  &   0.000958&0.000859  &  -0.000747&0.000451 & -0.00051&0.000427 \\
\multicolumn{9}{l}{{WRBE}}\\
\midrule 
$\mathbf{\gamma=0.2}$& -0.000480&0.000632  & -0.000540&0.000678  &  -0.000426&0.000584 &  -0.000231&0.000482 \\
$\mathbf{\gamma=0.4}$&0.000578&0.000876  &  -0.000708&0.000811  &   0.000581&0.000704 & -0.000012&0.000400 \\
$\mathbf{\gamma=0.6}$&0.000003&0.000401 &   0.000006 & 0.000395& -0.000142&0.000419&   0.000111& 0.000408  \\
$\mathbf{\gamma=0.8}$ &-0.000516 &0.000652 & -0.000138&0.000434 &  -0.000647&0.000765 &   0.000358&0.000951   \\
$\mathbf{\gamma=1.0}$&0.000354&0.000538 &  -0.000470& 0.000622&  -0.000093& 0.000424   &   0.000601&0.000725\\
\midrule
\multicolumn{9}{c}{{Dirichlet Prior}}\\
\midrule
{BE} &  0.000931&0.001003 &  -0.000935& 0.001018  & -0.000532 &0.000430  & -0.000813& 0.000557 \\
\midrule
\multicolumn{9}{l}{{WRBE}}\\
\midrule
$\mathbf{\gamma=0.2}$ &  0.000062 &0.000413&0.000163&0.000438&  0.000050& 0.000429  &  0.000453&0.000458 \\
$\mathbf{\gamma=0.4}$& 0.000420&0.000585  &   0.000458&0.000618 &  -0.0002567& 0.000477 &   0.000505&0.000253   \\
$\mathbf{\gamma=0.6}$ &  0.000415&0.000575  &  -0.000443&0.000601 &  -0.000628&0.000457 &   0.000191& 0.000476\\
$\mathbf{\gamma=0.8}$& -0.000222&0.000451  & -0.000121& 0.000402 &  -0.000369&0.000541  &   0.000248&0.000436\\
$\mathbf{\gamma=1.0}$ &  0.000464&0.000604  &  0.000314&0.000501 & -0.000051&0.000403 &   0.000311&0.000511 \\\bottomrule
\end{tabular}}}}
\label{tab8}
\end{table}
\subsection{Testing of Hypothesis Based on Robust Bayes Factor}
For the given data in Table \eqref{tab5}, testing of the hypothesis based on the robust Bayes factor is conducted here.  Let us consider a simple null hypothesis against the alternative as follows.
$$ 	\bm{H}_0 : \bm{\Lambda}=\bm{\Lambda}_0\quad \text{against}\quad \bm{H}_1 : \bm{\Lambda}\neq \bm{\Lambda}_0,$$
A continuous prior density would lead to zero prior probability to test $\bm{H}_0$.  Therefore, it is suggestive to take an $\epsilon$-neighbourhood (spherical) around $\bm{\Lambda}_0$.  The empirical prior and posterior probabilities are calculated to obtain the empirical Bayes factor.  From equation \eqref{bf}, the Bayes factor can be calculated using relation $$\text{Posterior odds}=\text{Prior odds}\times \text{Bayes factor}.$$  Here, the simple null hypothesis is taken as $\bm{\Lambda}_0=(-0.6, 0.34, 0.5, 0.1)$ and $\varepsilon=0.001$.  Table \ref{tab9} shows the empirically calculated Bayes factor for different tuning parameters under Normal and Dirichlet prior. 
\begin{table}[htb!]
\tbl{Empirical value of Bayes factor.}
{\scalebox{1}{
{\begin{tabular}{cccc} \toprule
{Tuning}& {Prior}&{Posterior}&{Bayes Factor}\\
{Parameter}&{odds}&{odds}& {BF$_{01}$} \\ \midrule
\multicolumn{4}{c}{{Normal prior}}\\
\midrule
$\bm{0.2}$&\multirow{5}*{0.162790}  &3.827585&23.51231\\
$\bm{0.4}$& &5.086957&31.24845\\
$\bm{0.6}$& &5.363635&32.94804\\
$\bm{0.8}$& &4.882353&29.99160\\
$\bm{1.0}$& &4.223880&25.94669\\\midrule
\multicolumn{4}{c}{{Dirichlet prior}}\\
\midrule
$\bm{0.2}$&\multirow{5}*{0.147727} &5.733331&38.81023\\
$\bm{0.4}$& &3.929579& 26.60022\\
$\bm{0.6}$& &4.223880&28.59241\\
$\bm{0.8}$& &4.263158&28.85829\\
$\bm{1.0}$& &6.769229&45.82246\\
\bottomrule
\end{tabular}}}}
\label{tab9}
\end{table}
Since the Bayes factor measures the strength of the evidence the data offers supporting one hypothesis over another, Jeffreys \cite{jeffreys1998theory} suggested a scale to interpret the Bayes factor and Kass and Raftery \cite{kass1995bayes} simplified it further, which is given in Table \ref{tab10}.  Observing table \ref{tab9} shows that support for $\bm{H}_0$ is strong.

\begin{table}[htb!]
\tbl{Interpretation of Bayes factor \cite{kass1995bayes}.}
{\scalebox{1}{
{\begin{tabular}{ll} \toprule
\textbf{BF$_{01}$}&\textbf{Support for $\bm{H}_0$}  \\ \midrule
$<1$&Negative\\
1 to 3& Not worth more than a bare mention\\
3 to 20 & Positive\\
20 to 150 & Strong\\
$>150$ & Very Strong\\ \bottomrule
\end{tabular}}}}
\label{tab10}
\end{table}
\subsection{Optimal Choice of Tuning Parameter}
As discussed in the introduction, the DPD measure-based estimation depends on the choice of tuning parameter $\gamma$.  Hence, finding the optimal value for tuning the parameter concerning the interest criteria is required. Here, We suggest a non-iterative method based on the approach introduced by Warwick and Jones \cite{w2005}, which involves minimizing the objective function
\begin{equation}
\Phi_{\gamma}(\hat{\bm{\Lambda}}){=}C_1\,D^w_{\gamma}({\hat{\bm{\Lambda}}})+C_2\,tr\Big(Q_{\gamma}^{-1}({\bm{\Lambda}})R_{\gamma}({\bm{\Lambda}})Q_{\gamma}^{-1}({\bm{\Lambda}})\Big), \label{tune}
\end{equation}
where, $C_1, C_2$ are predefined positive weight values with $C_1+C_2=1$.  For the given data in Table \eqref{tab5}, under different values of $\gamma$, WMDPDE are obtained and $\Phi_{\gamma}(\hat{\bm{\Lambda}})$ is calculated with $C_1=C_2=0.5.$  From the results presented in Table \ref{tab11}, $\gamma=0.75$ is the optimal value of the tuning parameter in this investigated case.
\begin{table}[htb!]
\tbl{Optimal value of tuning parameter $\gamma$.}
{\scalebox{1}{
{\begin{tabular}{lccc} \toprule
$\bm{\gamma}$&$\Phi_{\gamma}(\hat{\bm{\Lambda}})$&$\bm{\gamma}$&$\Phi_{\gamma}(\hat{\bm{\Lambda}})$\\ \midrule
0.10& 0.3759697&0.15& 0.3688173\\
0.20 &0.4059839&0.25& 0.5476472\\
0.30& 0.2162255&0.35 &0.2020160\\
0.40 &0.1891334&0.45 &0.1775084\\
0.50 &0.1671194&0.55 &0.1580054\\
0.60& 0.1502879&0.65& 0.1442069\\
0.70 &0.1401766&$\bm{0.75}$ &$\bm{0.1388715}$\\
0.80& 0.1413516&0.85& 0.1492402\\
0.90& 0.1649594&0.95& 0.1920193\\
1.00& 0.2353263\\ \bottomrule
\end{tabular}}}}
\label{tab11}
\end{table}
\section{Conclusion}\label{sec7}
This work is focused on studying robust estimation in the Bayesian framework under two competing causes of failures formulated by Lindley's lifetime distribution in the context of one-shot device data analysis where the robustified posterior density is developed on the exponential form of the maximizer equation based on the density power divergence method.  Through extensive simulation experiments,  the robustness of the weighted minimum density power divergence estimator (WMDPDE) and weighted robust Bayes estimator (WRBE) has been proved over the conventional maximum likelihood estimator (MLE) and Bayesian estimator (BE), respectively.  It has also been found that when data is contaminated, the bias of the WRBE is the least among the four estimation methods.  However, when prior information cannot be obtained, one can rely on a WMDPDE.  Robust hypothesis testing based on the Bayes factor has also been studied.  Further, the influence function, which measures the robustness analytically, has been derived and is shown graphically. 
 Finally, a pancreatic cancer dataset has been taken for real-life lifetime data analysis to establish the utility of the theoretical results explained in this work.

The model analyzed here can be implemented under step stress, assuming other lifetime distributions.  This study can be extended to the situation of dependent competing risks.  The missing cause of failure analysis can also be conducted.  Efforts in this direction are in the pipeline, and we are optimistic about reporting these findings soon.

\section*{Disclosure statement}
The authors report there are no competing interests to declare.

\section*{Data availability statement}
Surveillance, Epidemiology, and End Results (SEER) Program (www.seer.cancer.gov) SEER*Stat Database: Incidence - SEER Research Data, 8 Registries, Nov 2021 Sub (1975-2019) - Linked To County Attributes - Time Dependent (1990-2019) Income/Rurality, 1969-2020 Counties, National Cancer Institute, DCCPS, Surveillance Research Program, released April 2022, based on the November 2021 submission.

\bibliographystyle{apalike}
\bibliography{interacttfssample}

\section{Appendices}
\appendix

\section{Proof of Result \eqref{thm1}}

We denote failure probabilities and survival probability for $i^{th}$ group as
\begin{multicols}{4}
	\noindent $p_{i11}=P_{i1}$\\
	$p_{i21}=P_{i2}$\\
	$p_{i12}=P_{i3}$\\
	$p_{i22}=P_{i4}$\\
	$\dots\;\dots$\\
	$p_{il1}=P_{i(M-2)}$\\
	$p_{il2}=P_{i(M-1)}$\\
	$p_{i0}=P_{iM}.$
\end{multicols}
\noindent Thus, $h=2(l-1)+r$ and $M=2L+1$ i.e. $p_{ilr}=P_{i[2(l-1)+r]}$.  For the $i^{th}$ group, the number of failures between the interval $(\tau_{i(l-1)}-\tau_{il})$ is denoted as, $N_{il}=n_{il1}+n_{il2}$.  Again, we define, $X_{ui}=(X_{ui1},X_{ui2},\dots,X_{uiM})\sim MN(1,\undertilde{P_i})$; where, $\undertilde{P_i}=(P_{i1},P_{i2},\dots,P_{iM})$.  Therefore, $N_{ih}=\sum_{u_i=1}^{g_i}X_{u_{i}h}$.  Hence, the WDPD measure in equation \eqref{dpd} ignoring the terms independent of parameters is given as
\begin{equation*}
H_g(\gamma)=\sum_{i=1}^{I}\frac{g_i}{G}\left[\sum_{h=1}^{M}(P_{ih})^{\gamma+1}-\frac{\gamma+1}{\gamma}\sum_{h=1}^{M}\frac{N_{ih}}{g_i}(P_{ih})^{\gamma}\right]. 
\end{equation*}
The proof of the theorem proceeds based on Calvino et al. \cite{cal2021}.  The simplified version of the proof in the present context is given in the study of Baghel and Mondal \cite{baghel2024analysis},
\begin{flalign*}
Q_{\gamma}(\bm{\Lambda})&=\left[\left(\sum_{i=1}^{I}\frac{g_i}{G}\sum_{h=1}^{M}P_{ih}^{\gamma-1}\frac{\partial(P_{ih})}{\partial\bm{\Lambda}_j}\frac{\partial(P_{ih})}{\partial\bm{\Lambda}_f}\right)_{j,f}\right]\;;\;(j,f){=}1,...,4.\\
R_{\gamma}(\bm{\Lambda})&=\left[\left\{\sum_{i=1}^{I}\frac{g_i}{G}\left(\sum_{h=1}^{M}P_{ih}^{2\gamma-1}(1-P_{ih})\frac{\partial(P_{ih_1})}{\partial\bm{\Lambda}_j}\frac{\partial(P_{ih_2})}{\partial\bm{\Lambda}_f}{-}2\sum_{(h_1,h_2)}P_{ih_1}^{\gamma}P_{ih_2}^{\gamma}\frac{\partial(P_{ih_1})}{\partial\bm{\Lambda}_j}\frac{\partial(P_{ih_2})}{\partial\bm{\Lambda}_f}\right)\right\}_{j,f}\right].
&&
\end{flalign*}

\section{Proof of Result \ref{res4}}
Let us denote
\begin{flalign*}
T^{(\gamma)}_G(F_{\bm{\Lambda}})=\frac{\int \bm{\Lambda}\exp\left\{ B^w_{\gamma}(\bm{\Lambda};F_{\bm{\Lambda}})\right\}\pi(\bm{\Lambda})d\bm{\Lambda}}{\int\exp\left\{B^w_{\gamma}(\bm{\Lambda};F_{\bm{\Lambda}})\right\}\pi(\bm{\Lambda})d\bm{\Lambda}}=\frac{A_1(\bm{\Lambda};F_{\bm{\Lambda}})}{A_2(\bm{\Lambda};F_{\bm{\Lambda}})}.
\end{flalign*}
Then, the IF of WRBE can be obtained as
\begin{flalign*}
IF(t;T^{(\gamma)}_G,F_{\bm{\Lambda}})&=\left.\frac{\partial}{\partial\epsilon}T^{(\gamma)}_G(M_{\epsilon})\right\vert_{\epsilon\to 0^{+}}=\frac{\int \bm{\Lambda}X_{\gamma}(\bm{\Lambda};\bm{t},f_{\bm{\Lambda}})\exp\left\{B^w_{\gamma}(\bm{\Lambda})\right\}\pi(\bm{\Lambda})d\bm{\Lambda}}{\int\exp\left\{B^w_{\gamma}(\bm{\Lambda})\right\}\pi(\bm{\Lambda})d\bm{\Lambda}}\\
&\quad\;-\left[\frac{\int \bm{\Lambda}\exp\left\{ B^w_{\gamma}(\bm{\Lambda})\right\}\pi(\bm{\Lambda})d\bm{\Lambda}}{\int\exp\left\{B^w_{\gamma}(\bm{\Lambda})\right\}\pi(\bm{\Lambda})d\bm{\Lambda}}\right.\left.\times\frac{\int X_{\gamma}(\bm{\Lambda};\bm{t},f_{\bm{\Lambda}})\exp\left\{B^w_{\gamma}(\bm{\Lambda})\right\}\pi(\bm{\Lambda})d\bm{\Lambda}}{\int\exp\left\{B^w_{\gamma}(\bm{\Lambda})\right\}\pi(\bm{\Lambda})d\bm{\Lambda}}\right]\\
&\quad=Cov_{(p)}\left(\bm{\Lambda},X_{\gamma}(\bm{\Lambda};\bm{t},f_{\bm{\Lambda}})\right),
	&&
\end{flalign*}
where, $Cov_{(p)}()$ is the covariance for posterior distribution.
\section{Proof of Result \ref{res5}}
Let us denote
\begin{flalign*}
&\frac{\int_{\bm{\Theta}_0}\exp\big\{B^w_{\gamma}(\bm{\Lambda}\in\bm{\Theta}_0;F_{\bm{\Lambda}_0})\big\}\pi_0(\bm{\Lambda})\,d\bm{\Lambda}}{\int_{\bm{\Theta}_1}\exp\big\{B^w_{\gamma}(\bm{\Lambda}\in\bm{\Theta}_1;F_{\bm{\Lambda}_0})\big\}\pi_1(\bm{\Lambda})\,d\bm{\Lambda}}=\frac{R_0(\bm{\Lambda}\in\bm{\Theta}_0)}{R_1(\bm{\Lambda}\in\bm{\Theta}_1)}.
\end{flalign*}
Then, the Influence function of the Bayes factor can be obtained as
\begin{flalign*}
IF(t;T^{(\gamma)}_{\bm{\Theta}},F_{\bm{\Lambda}_0})&=\left.\frac{\partial(T^{(\gamma)}_{\bm{\Theta}}(M_{\epsilon}))}{\partial\epsilon}\right\vert_{\epsilon\to 0^{+}}\\
&=\frac{1}{\left\{R_1(\bm{\Lambda}\in\bm{\Theta}_1)\right\}^2}\left[R_1(\bm{\Lambda}\in\bm{\Theta}_1)\frac{\partial}{\partial\epsilon}R_0(\bm{\Lambda}\in\bm{\Theta}_0)\right.\\
&\qquad-\left.\left.R_0(\bm{\Lambda}\in\bm{\Theta}_0)\frac{\partial}{\partial\epsilon}R_1(\bm{\Lambda}\in\bm{\Theta}_1)\right]\right\vert_{\epsilon\to 0^{+}}\\
&{=}\left[\frac{\int_{\bm{\Theta}_0}X_{\gamma}(\bm{\Lambda}\in\bm{\Theta}_0)\exp\big\{B^w_{\gamma}(\bm{\Lambda}\in\bm{\Theta}_0)\big\}\pi_0(\bm{\Lambda})\,d\bm{\Lambda}}{\int_{\bm{\Theta}_0}\exp\big\{B^w_{\gamma}(\bm{\Lambda}\in\bm{\Theta}_0)\big\}\pi_0(\bm{\Lambda})\,d\bm{\Lambda}}{\times}Y_{\gamma}(\bm{\Theta})\right]\\
&\quad{-}\left[Y_{\gamma}(\bm{\Theta}){\times}\frac{\int_{\bm{\Theta}_1}X_{\gamma}(\bm{\Lambda}\in\bm{\Theta}_1)\exp\big\{B^w_{\gamma}(\bm{\Lambda}\in\bm{\Theta}_1)\big\}\pi_1(\bm{\Lambda})\,d\bm{\Lambda}}{\int_{\bm{\Theta}_1}\exp\big\{B^w_{\gamma}(\bm{\Lambda}\in\bm{\Theta}_1)\big\}\pi_1(\bm{\Lambda})\,d\bm{\Lambda}}\right]\\
&=Y_{\gamma}(\bm{\Theta})\bigg\{E\Big[X_{\gamma}(\bm{\Lambda}\in\bm{\Theta}_0)\Big]-E\Big[X_{\gamma}(\bm{\Lambda}\in\bm{\Theta}_1)\Big]\bigg\},
&&
\end{flalign*}
where,
\begin{flalign*}
X_{\gamma}(\bm{\Lambda}\in\bm{\Theta}_j)&=\left.\frac{\partial(B^w_{\gamma}(\bm{\Lambda}\in\bm{\Theta}_j;F_{\bm{\Lambda}_0}))}{\partial\epsilon}\right\vert_{\epsilon\to 0^{+}}\;;\;j=0,1.\\
&=\frac{1}{\gamma}\sum_{i=1}^{I}\frac{g_i}{G}\Bigg[\Big\{\delta_{I_{i0}}(t)-p_{i0}(\bm{\Lambda}_0)\Big\}p^{\gamma}_{i0}(\bm{\Lambda}\in\bm{\Theta}_j)\Bigg.\left.+\sum_{l=1}^{L}\sum_{r=1}^{2}\Big\{\delta_{I_{ilr}}(t)-p_{ilr}(\bm{\Lambda}_0)\Big\}p^{\gamma}_{ilr}(\bm{\Lambda}\in\bm{\Theta}_j)\right].
&&
\end{flalign*}
\end{document}